\documentclass[]{JHEP3}

\usepackage{epsfig,multicol,bbm,graphicx}
\usepackage{graphicx}
\usepackage{epstopdf}
\usepackage{latexsym}

\newcommand\fverb{\setbox\fverbbox=\hbox\bgroup\verb}
\newcommand\fverbdo{\egroup\medskip\noindent%
            \fbox{\unhbox\fverbbox}\ }
\newcommand\fverbit{\egroup\item[\fbox{\unhbox\fverbbox}]}
\newbox\fverbbox

\newcommand{\la}{\lambda}

\newcommand{\bear}{\begin{eqnarray}}
\newcommand{\eear}{\end{eqnarray}}

\newbox\pippobox

\def\be{\begin{equation}}
\def\ee{\end{equation}}
\def\bea{\begin{eqnarray}}
\def\eea{\end{eqnarray}}
\def\bx{{\bf x}}

\def\bk{{\bf k}}
\def\bp{{\bf p}}
\def\bn{{\bf n}}

\def\a{\alpha}

\def\m{\mu}
\def\n{\nu}
\def\9{\nabla}
\def\r{\rho}
\def\s{\sigma}
\def\vs{\varsigma}

\def\T{\Theta}
\def\g{\gamma}
\def\z{\zeta}
\def\b{\beta}
\def\o{\omega}
\def\lam{\lambda}

\def\kp{\kappa}
\def\h{\eta}
\def\d{\delta}
\def\dd{{\rm d}}
\def\D{\Delta}

\def\nn{\nonumber}
\def\half{\frac12}
\def\le{\left}
\def\ri{\right}
\def\6{\partial}

\def\ve{\varepsilon}
\def\f{\frac}
\def\ma{\mathcal}
\def\la{\langle}
\def\ra{\rangle}

\def\Mp{M_{pl}}
\def\tld{\tilde}
\def\0{(0)}
\def\half{\f{1}{2}}
\def\>{\rightarrow}

\title{Acoustic signatures in the Cosmic Microwave Background
    bispectrum from primordial magnetic fields}

\author{Rong-Gen Cai, Bin Hu, Hong-Bo Zhang\\
    Key Laboratory of Frontiers in Theoretical Physics,
    Institute of Theoretical Physics, Chinese Academy of Sciences,
    P.O. Box 2735, Beijing 100190, China\\
    E-mail:
    \email{cairg@itp.ac.cn}
    \email{hubin@itp.ac.cn}
        \email{hbzhang@itp.ac.cn}}

\received{\today}       
\accepted{\today}       

\abstract{ Using the full radiation transfer function, we
numerically calculate the CMB angular bispectrum seeded by the
compensated magnetic scalar density mode. We find that, for the
string inspired primordial magnetic fields characterized by index
$n_B=-2.9$ and mean-field amplitude $B_{\lam}=9{\rm~nG}$, the
angular bispectrum is dominated by two primordial magnetic shapes.
The first magnetic shape looks similar to the one from local-type
primordial curvature perturbations, so both the amplitude and
profile of the Komatsu-Spergel estimator (reduced bispectrum)
seeded by this shape are almost the same as those of the primary
CMB anisotropies. However, for different parameter sets
($l_1,l_2$), this ``local-type'' reduced bispectrum oscillates
around different asymptotic values in the high-$l_3$ regime
because of the effect of the Lorentz force, which is exerted by
the primordial magnetic fields on the charged baryons. This
feature is  different from the standard case where all modes
approach to zero asymptotically in the high-$l$ limit. On the
other hand, the second magnetic shape  appears only in the
primordial magnetic field model.  The amplitude of the
Komatsu-Spergel estimator sourced by the second shape diverges in
the low-$l$ regime because of the negative slope of shape. In the
high-$l$ regime, this amplitude is approximately equal to that of
the first estimator, but with a reversal phase.}

\begin{document}


\section{\label{intro}Introduction}
In the inflationary scenario~\cite{Sato:1980yn}, the quantum
fluctuations of the scalar field(s) are responsible to generate
the initial conditions for the Cosmic Microwave Background (CMB)
anisotropies. The current observations \cite{Komatsu:2010fb} from
large scale structures are consistent with an almost scale
invariant, Gaussian primordial density perturbations generated
during inflation. However, with the improvements of measurement
precision, any small deviations from the Gaussian distribution
enable us to distinguish different cosmological models. Like the
role colliders play in particle physics, measurements of
non-Gaussian features provide microscopic information on the
interactions of the inflatons and/or curvatons. Constraining and
detecting non-Gaussianity (NG) have become one of the major
efforts in modern cosmology. A variety of potentially detectable
forms of primordial non-Gaussian features from inflation models
have been intensively investigated (see
\cite{Bartolo:2004if,Chen:2010xk} for a review). The effects of
primordial non-Gaussian curvature perturbations on CMB
anisotropies have also been studied in the recent papers
\cite{Yadav:2010fz,Liguori:2010hx,Bartolo:2010qu,Komatsu:2010hc,
Fergusson:2009nv,Fergusson:2010dm,Smidt:2010ra,Pitrou:2010sn,
Nitta:2009jp,Pitrou:2008ak,Fergusson:2008ra}. The current
limitations on the primordial bispectra from WMAP-$7$yr data are
$-10<f_{{\rm NL}}^{{\rm local}}<74$ and $-214<f_{{\rm NL}}^{{\rm
equil}}<266$ at $95\%$ CL \cite{Komatsu:2010fb,Komatsu:2010hc},
where $f_{{\rm NL}}^{{\rm local}}$ and $f_{{\rm NL}}^{{\rm
equil}}$ are the non-linear parameters of the ``squeezed" and
``equilateral" momentum configurations, respectively.

Except for the possible NG from the inflationary dynamics, the
primordial NG might come from other mechanisms. One interesting
possibility is that the non-Gaussianities are sourced by the
primordial magnetic fields (PMFs) in the large scale structures
\cite{Brown:2005kr,Seshadri:2009sy,Caprini:2009vk}. The
astrophysical observations about the spiral/elliptical galaxies
and rich clusters indicate that our universe is permeated with
large scale coherent magnetic fields with the magnitudes ranging
from hundreds of ${\rm nG}$ to few $\m{\rm G}$
\cite{Carilli:2001hj,Beck:1995zs,Xu:2005rb,Kronberg:1993vk},
however, their origins are still not yet fully understood. The
dynamo mechanism explains the origin of the galactic magnetic
fields with amplification of a small frozen-in seed field to the
observed $\m{\rm G}$ field through turbulence and differential
rotation \cite{zeldovich83,parker79}. And the gravitational
adiabatic compression may generate the magnetic fields in clusters
during the collapse of a protogalactic cloud
\cite{piddington64,ohki64,kulsrud90}. Cosmological phase
transitions in the early universe may produce the tiny magnetic
seed fields, which are required by the above mentioned
amplification mechanisms, such as the electroweak phase transition
\cite{Baym:1995fk,Sigl:1996dm}, QCD phase transition
\cite{Quashnock:1988vs,Cheng:1994yr} and the inflation with the
broken conformal invariance \cite{Turner:1987bw}.

In recent years, intensive effort has been devoted to studying the
imprints of magnetic fields on the CMB anisotropies, which are
nicely reviewed in \cite{Giovannini:2003yn}. The contributions to
the CMB angular power spectrum from the scalar perturbations induced
by PMFs are investigated in
\cite{Koh:2000qw,Giovannini:2007ks,Giovannini:2004aw,
scalarGiovannini, Giovannini:2007qn,Kahniashvili:2006hy,
Ichiki:2006cd,Yamazaki:2008gr,Paoletti:2008ck,
Kojima:2009ms,Shaw:2009nf,Finelli:2008xh,Bonvin:2010nr}, from the
vector perturbations in
\cite{Lewis:2004ef,Durrer:1998ya,Mack:2001gc,Yamazaki:2008gr,Paoletti:2008ck} and
from tensor mode in
\cite{Mack:2001gc,Durrer:1999bk,Caprini:2003vc,Yamazaki:2008gr,Paoletti:2008ck},
respectively. Some other phenomena induced by PMFs, such as Faraday
rotation, damping of Aflv\'{e}n waves, effects of PMFs on seeds for
large scale structures and on neutrino masses are investigated in
\cite{Kosowsky:2004zh}, \cite{Jedamzik:1996wp,Subramanian:1997gi},
\cite{Yamazaki:2006mi} and \cite{Kojima:2008rf,Yamazaki:2010jw},
respectively. And the new constraints on PMFs from CMB anisotropy
and large scale structure data are reported in
\cite{Yamazaki:2010nf,Yamazaki:2008jh,Paoletti:2010rx}.

In the inflationary scenario, the NG signals come from the high
order curvature perturbations. However, even at the lowest order,
PMFs can  still generate some non-Gaussian features in the CMB
anisotropies, since the magnetic energy density and anisotropic
stress induced by PMFs are naturally non-Gaussian variables. Such
signatures have been investigated in
\cite{Chen:2004nf,Naselsky:2004gm,Bernui:2008ve,
Kahniashvili:2008sh,Durrer:1998ya,Naselsky:2008ei,
Demianski:2007fz,Bernui:2008cr}, but for the homogeneous magnetic
fields with fixed direction which break the spatial isotropy and
result in the north-south asymmetry on the CMB sky. However, as
pointed out in \cite{Seshadri:2009sy,Caprini:2009vk}, the
stochastic PMFs are able to generate a distinctive non-Gaussian
signal in the CMB anisotropies with an amplitude comparable with
the one from the primary curvature perturbations.

The authors in \cite{Seshadri:2009sy,Caprini:2009vk} analytically
calculate the CMB bispectrum from the stochastic PMFs, but only in
the Sachs-Wolfe regime ($l\leq10$). In this paper we calculate the
angular bispectrum from scalar perturbations induced by PMFs with
the full transfer function. Because the scale-invariant magnetic
power spectra are strongly inspired by string cosmological model
as a consequence of the breaking of conformal invariance during
the pre-big bang phase \cite{Gasperini:1995dh,Gasperini:1995fm}, following
most of the literatures in this subject (for example, $n_B\simeq-3$ in
\cite{Seshadri:2009sy} and $n_B=-2.9,\pm2$ in \cite{Caprini:2009vk}), in
this paper we take the magnetic index $n_B=-2.9$ and mean-field
amplitude $B_{\lam}=9{\rm~nG}$. In this model, we find that the
angular bispectrum is dominated by two primordial magnetic shapes.
The first magnetic shape $f^{(1)}(k,q,p)$ (\ref{f1}) looks similar
to the one from local-type primordial curvature perturbations, so
both the amplitude and profile of the Komatsu-Spergel estimator
(reduced bispectrum) seeded by this shape are almost the same as
those of the primary CMB anisotropies
\cite{Komatsu:2001rj,Babich:2004yc}, (see Figure \ref{ba},
\ref{bc}, \ref{lll1} and \ref{3l1}). However, for different
parameter sets ($l_1,l_2$), this ``local-type'' estimator
$b^{(1)}_{l_1l_2l_3}$ (\ref{blll1}) oscillates around different
asymptotic values in the high-$l_3$ regime because of the effect
of the Lorentz force, (see Figure \ref{3l1} and
\ref{nolorentz3l1}). This feature is  different from the standard
case where all modes approach to zero asymptotically in the
high-$l$ limit. On the other hand, the second magnetic shape
$f^{(2)}(k,q,p)$ (\ref{f2}) appears only in the primordial
magnetic field model. However, the amplitude of the
Komatsu-Spergel estimator $b^{(2)}_{l_1l_2l_3}$ (\ref{blll2})
sourced by the shape $f^{(2)}(k,q,p)$ diverges in the low-$l$
regime because of the negative slope of shape. In the high-$l$
regime, this amplitude is approximately equal to that of the first
estimator $b^{(1)}_{l_1l_2l_3}$, but with a reversal phase, (see
Figure \ref{bd}, \ref{lll2} and \ref{3l2}).

The rest of this paper is organized as follows. In section
\ref{spmf}, we firstly present the Maxwell and conservation
equations which govern the behaviors of electromagnetic fields in
the curved spacetime. Then we calculate the primordial magnetic
power spectrum induced by PMFs under the ideal
magnetohydrodynamics approximation. The linearized scalar
equations for each individual matter component in the Cold Dark
Matter (CDM) model and the gravitational fields are given in
section \ref{eq}. In section \ref{ic}, we derive two magnetic
initial conditions in the deep radiation dominant era, and then
calculate the CMB angular power spectrum numerically by using
these initial conditions. The numerical calculations about CMB
bispectrum signatures seeded by the compensated magnetic density
mode are analyzed in section \ref{bispectrum}. Finally, we
conclude in section \ref{concl}.

\section{\label{spmf}Stochastic primordial magnetic fields}
In this section, we firstly present the Maxwell and conservation
equations which govern the evolution of electromagnetic fields in
a curved spacetime. Then we calculate the primordial magnetic
power spectrum induced by PMFs under the ideal
magnetohydrodynamics approximation.

\subsection{Electromagnetic field in a curved spacetime}
In this subsection, we present the Maxwell's equations and
conservation equations in a covariant formulism\footnote{The
covariant approach to cosmological perturbations is shortly
reviewed in  Appendix \ref{cov}, and the definitions of covariant
variables, such as 4-velocity $u^a$, expansion rate $\T$, etc. can
be found there.}. Following the formulism, the electromagnetic
(Faraday) tensor $F_{ab}$ can be decomposed into an electric and a
magnetic component as
 \be\label{fara}
 F_{ab}=2u_{[a}E_{b]}+\ve_{abc}B^c\;,\ee
where $E_a=F_{ab}u^b$ and $B_a=\ve_{abc}F^{bc}/2$ are respectively
the electric and magnetic fields experienced by the observer with
4-velocity $u^a$ ($E_au^a=B_au^a=0$). The Faraday tensor also
determines the  energy-momentum tensor of the electromagnetic
field as\footnote{ In this paper we take the unit conventions as
$c=\hbar=\Mp=1/8\pi{\rm G}=1$.}
 \be\label{em tensor}
 T_{ab}^{(em)}=\f{1}{4\pi}\le[-F_{ac}F^c_{~b}-\f{1}{4}F_{cd}F^{cd}g_{ab}\ri]\;.\ee
Combining (\ref{fara}) with (\ref{em tensor}), we arrive at the irreducible form of $T_{ab}^{(em)}$
 \be\label{em tensor2}
  T_{ab}^{(em)}=\f{1}{4\pi}\le[\f{1}{2}(E^2+B^2)u_au_b+\f{1}{6}(E^2+B^2)h_{ab}+2q_{(a}u_{b)}\ri]+\pi^{(B)}_{ab}\;.\ee
Here $E^2=E_aE^a$ and $B^2=B_aB^a$ are the square magnitudes, $q_a=\ve_{abc}E^bB^c$ and
$\pi^{(B)}_{ab}=(-E_{\la
a}E_{b\ra}-B_{\la a}B_{b\ra})/4\pi$ are the
electromagnetic Poynting vector and
anisotropic stress tensor, respectively \footnote{Comparing with
the convensional definition about the electromagnetic anisotropic
tensor $\pi^{(B)}_{ab}$, such as the one in the Jackson's textbook
\cite{jackson75}, the definition in this paper is different from
the convensional one by a factor of $1/4\pi$.}.
In this paper, the round, squared and angled brackets denote the
 symmetric, anti-symmetric, and symmetric trace-free parts of a
 tensor, respectively.

In the standard tensor form the Maxwell equations read
 \be\label{Maxwell}
 \9_bF^{ab}=J^a\;,\qquad \9_{[c}F_{ab]}=0\Longleftrightarrow \h^{abcd}F_{bc;d}=0\;,\ee
where $J^a$ is the 4-current that sources the electromagnetic field.
With respect to the $u_a$-congruence, the 4-current splits into its
irreducible parts according to
 \be\label{current}
 J^a=\m u^a+\ma{J}^a\;,\ee
with $\m=-J_au^a$, $\ma{J}^a=h^a_{~b}J^b$ and $\ma{J}_au^a=0$.
By virtue of the irreducible form of $F_{ab}$ and $J^a$,
the timelike parts of the Maxwell equations read
 \bea
 h_a^{~c}\dot E_c&=&-\f{2}{3}\Theta E_a+(\s_{ab}+\ve_{abc}\o^c)E^b+\ve_{abc}A^bB^c
 +{\rm curl}~B_a-\ma{J}_a\;,\label{Max1}\\
 h_a^{~c}\dot B_c&=&-\f{2}{3}\Theta B_a+(\s_{ab}+\ve_{abc}\o^c)B^b-\ve_{abc}A^bE^c
 -{\rm curl}~E_a\;,\label{Max2}\eea
while their spacelike components provide the constraints
 \bea
 D_aE^a+2\o_aB^a&=&\m\;,\label{Max3}\\
 D_aB^a-2\o_aE^a&=&0\;,\label{Max4}\eea
where $D_a$ denotes the spatial derivatives with respect to
projected metric $h_{ab}$ and its definition is presented
in (\ref{deriv}).

Besides, the 4-current conservation law $\9_aJ^a=0$ gives the
continuity equation of charge density
 \be\label{charge consv}
 \dot\m=-\Theta\m-D_a\ma{J}^a-A_a\ma{J}^a\;.\ee
The equations
(\ref{Max1}),(\ref{Max2}),(\ref{Max3}),(\ref{Max4}),(\ref{charge
consv}) form a complete set of equations which evolve the
electromagnetic field in a curved spacetime.

\subsection{Ideal MHD approximation in the Universe}
A good conductor throughout the history of the Universe allows us to
study the electromagnetic field in the universe within the limits of
ideal magnetohydrodynamics (MHD) approximation. By means of Ohm's
law, the spatial currents $\ma{J}_a$ read
 \be\label{Ohm}
 \ma{J}_a=\vs E_a\;,\ee
where $\vs$ represents the scalar conductivity of the medium. The
MHD approximation states that, in
the limit $\vs\rightarrow\infty$, we can neglect the electric field
$E_a$. Hence, the energy-momentum tensor of the residual magnetic field becomes
 \be\label{magTab}
 T_{ab}^{(B)}=\f{1}{4\pi}\le[\f{1}{2}B^2u_au_b+\f{1}{6}B^2h_{ab}\ri]+\pi^{(B)}_{ab}\;,\ee
with the anisotropic tensor $\pi^{(B)}_{ab}=-B_{\la
a}B_{b\ra}/4\pi$. From the above expression we can identify the
energy density of PMFs as $\D^{(B)}=B^2/8\pi$. In addition, the
Maxwell equations reduce into a single propagation equation
 \bea\label{Max5}
 \dot B_{\la a\ra}&=&\le(\s_{ab}+\ve_{abc}\o^c-\f{2}{3}\T
 h_{ab}\ri)B^b\;,\eea
and three constraints
 \bea
 \ma{J}_a&=&{\rm curl}~B_a+\ve_{abc}A^bB^c\;,\\\label{Max6}
 \m&=&2\o^aB_a\;,\\\label{Max7}
 0&=&D^aB_a\;.\label{Max8}\eea

\subsection{Primordial power spectrum induced by PMFs}
In this subsection, we present the primordial power spectrum
induced by scalar perturbations from the stochastic PMFs. In the
local rest frame $u^a=(1,\vec{0})$, $B_au^a=0$ leads to a
vanishing temporal component of $B_a$ and then we have
 \bea
 B_a(t,\bx)&\rightarrow& B_i(t,\bx)\;,\\
 \pi^{(B)}_{ab}(t,\bx)&\rightarrow& \pi^{(B)}_{ij}(t,\bx)=
 \f{1}{4\pi}\le[\f{1}{3}B^k(t,\bx)B_k(t,\bx)\d_{ij}-B_i(t,\bx)B_j(t,\bx)\ri]\;,\\
 \D^{(B)}(t,\bx)&\rightarrow& \D^{(B)}(t,\bx)=\f{B^i(t,\bx)B_i(t,\bx)}{8\pi}\;.
\eea Furthermore, in the ideal MHD regime we can separate out the
time evolution of PMFs, $B_i(t,\bx)=B_i(\bx)/a^2$. Hence, in what
follows we concentrate on the time independent spatial component
$B_i(\bx)$ and take them as statistically homogeneous and
isotropic random fields. The transversal nature of PMFs leads to
 \be\label{pmfspec}
 \la B_i(\bk)B^{\ast}_j(\bk')\ra=(2\pi)^3\f{P_{ij}}{2}P^{(B)}(k)\d(\bk-\bk')\;,
 \qquad k<k_D\;,\ee
where $P_{ij}=\d_{ij}-\hat k_i\hat k_j$ is the projector onto the
transverse plane, $k_D$ is the wavenumber of damping scale and
$P^{(B)}(k)$ is the primordial magnetic power spectrum. For some
specific magnetogenesis models $P^{(B)}(k)$ takes the power law
form
 \be\label{P_B}
  P^{(B)}(k)=Ak^{n_B}\;.\ee
In the above expression, we have adopted the Fourier transform
convention as
 \bea\label{fourierconv}
 B_i(\bx)&=&\f{1}{(2\pi)^3}\int d^3k \tld B_i(\bk)e^{-i\bk\cdot\bx}\;,\\
 \tld B_i(\bk)&=&\int d^3x B_i(\bx)e^{i\bk\cdot\bx}\;.\eea
It is convenient to introduce the Fourier components of the PMF
energy density contrast $\D^{(B)}_k$ and scalar part of the
anisotropic stress tensor $\pi^{(B)}_k$ as
 \bea\label{fourierstress}
 \D^{(B)}(\bx)&=&\f{1}{(2\pi)^3}\int d^3k ~\D^{(B)}_ke^{-i\bk\cdot\bx}\;,\\
 \pi^{(B)i}_{~~~~~j}(\bx)&=&\f{1}{(2\pi)^3}\int d^3k ~\pi^{(B)}_k\le(\f{1}{3}\d^i_{~j}-\hat k^i\hat k_j\ri)
 e^{-i\bk\cdot\bx}\;.\eea
Thus, we obtain the expressions for $\D^{(B)}_k$ and $\pi^{(B)}_k$
from the momentum convolution
 \bea\label{convolution}
 \D^{(B)}_k&=&\f{1}{8\pi}\int \f{d^3 p}{(2\pi)^3}\tld B^i(\bp)\tld B_i(\bk-\bp)\;,\\
 \pi^{(B)}_k&=&\f{3}{8\pi}\int \f{d^3 p}{(2\pi)^3}\le[
 \hat k_i\tld B^i(\bp)\hat k^j\tld B_j(\bk-\bp)-\f{1}{3}\tld B^i(\bp)\tld B_i(\bk-\bp)\ri]\;.\eea
Since we are interested in the PMFs in the linear perturbation
regime, we therefore define the magnetic comoving mean-field
amplitude by smoothing over a Gaussian sphere of the comoving
radius $\lam=1~{\rm Mpc}$ ($f_k=e^{-\lam^2k^2/2}$) as
 \be\label{normal}
 \la B_i(\bx)B_i(\bx)\ra|_{\lam}=B^2_{\lam}\;.\ee
For the power law model (\ref{P_B}), $B_{\lam}^2$ can be given by
the Fourier transform of the product of the power spectrum
$P^{(B)}(k)$ and the square of the filter transform $f_k$,
 \be\label{normal2}
 B_{\lam}^2=\f{2}{(2\pi)^3}\int d^3k P^{(B)}(k)|f_k|^2
 \simeq\f{2A}{(2\pi)^2}\f{1}{\lam^{n_B+3}}\Gamma\le(\f{n_B+3}{2}\ri)\;,\ee
where we require the spectral index $n_B>-3$ to prevent the
infrared divergence at the power spectrum level. Plugging
(\ref{normal2}) into (\ref{pmfspec}), we arrive at
 \be\label{pmfspec2}
 \la B_i(\bk)B^{\ast}_j(\bk')\ra=(2\pi)^3\f{P_{ij}}{2}\f{(2\pi)^{n_B+5}B_{\lam}^2}{2\Gamma\le(\f{n_B+3}{2}\ri)}
 \f{k^{n_B}}{k_{\lam}^{n_B+3}}\d(\bk-\bk')\;,\qquad k<k_D\;,\ee
where $k_{\lam}=2\pi/\lam$.
For all scales smaller than the damping
scale ($k>k_D\simeq 4.5~{\rm Mpc}^{-1}$) the spectrum vanishes.

Furthermore, we can obtain the two-point correlation functions for $\D^{(B)}_k$ and $\pi^{(B)}_k$
by using the Wick theorem
 \bea
 \la \D^{(B)}(\bk)\D^{(B)\ast}(\bk')\ra&=&\f{\d(\bk-\bk')}{128\pi^2}\int d^3p~P^{(B)}(p)P^{(B)}(|\bk-\bp|)
 \le(1+\m^2\ri)\;,\label{DB_2pt}\\
 \la \pi^{(B)}(\bk)\pi^{(B)\ast}(\bk')\ra&=&\f{\d(\bk-\bk')}{32\pi^2}\int d^3p~P^{(B)}(p)P^{(B)}(|\bk-\bp|)\nn\\
 &&\le[1-\f{3}{4}(\g^2+\beta^2)+\f{9}{4}\g^2\beta^2-\f{3}{2}\g\beta\m+\f{1}{4}\m^2\ri]\;,\label{piB_2pt}
 \eea
where $\m=\hat{p}\cdot(\widehat{\bk-\bp})$,
$\g=\hat{k}\cdot\hat{p}$ and
$\beta=\hat{k}\cdot(\widehat{\bk-\bp})$. Following most of the
literatures in this subject, and also due to that the nearly
scale-invariant magnetic power spectra are strongly inspired by
the string cosmological models as a consequence of the breaking of
conformal invariance during the pre-big bang phase
\cite{Gasperini:1995dh,Gasperini:1995fm}, we take $n_B=-2.9$ in the following
calculations. Ignoring the cutoff in the definitions of $P^{(B)}$
allows us to integrate (\ref{DB_2pt}) and (\ref{piB_2pt})
semi-analytically \cite{Shaw:2009nf}
 \bea
 P_{\D_B}(k)&\simeq& \f{42.37}{16}
 \le[\f{(2\pi)^{n_B+2}B_{\lam}^2}{2\Gamma(\f{n_B+3}{2})\r^{(\g)}}\ri]^2
 \le(\f{k}{k_{\lam}}\ri)^{2n_B+6}\;,\label{pow_DB}\\
 P_{\pi_B}(k)&\simeq& \f{9\times 14.55}{4}
 \le[\f{(2\pi)^{n_B+2}B_{\lam}^2}{2\Gamma(\f{n_B+3}{2})\r^{(\g)}}\ri]^2
 \le(\f{k}{k_{\lam}}\ri)^{2n_B+6}\;,\label{pow_piB}\eea
where we have used the convention about dimensionless power
spectrum
 \be
 \la X(\bk)X^{\ast}(\bk')\ra=2\pi^2(2\pi)^3\d(\bk-\bk')k^{-3}P_{X}(k)\;,\qquad
 X=\D_B,~\pi_B\;.\ee

\section{\label{eq}Basic equations}
In this section we present the magnetic linearized scalar
equations in Fourier space\footnote{The set of the linear
equations for all matter components in the coordinate space can be
found in Appendix (\ref{linear_equations})}. Firstly, we define
the scalar-valued harmonic function on the exact
Friedmann-Robertson-Walker (FRW) background
 \be\label{scalar eigen}
 a^2D^2\ma{Q}^{(0)}(k)+k^2\ma{Q}^{(0)}(k)=0\;,\qquad
 \dot\ma{Q}^{\0}(k)=0\;,\ee
where $a$ is the scale factor and the superscript $(0)$ represents
 the scalar mode. The covariant temporal and spatial derivatives
are defined in (\ref{deriv}). Arming with the scalar harmonics, we
can calculate the rank-$l$ Projected Symmetric and Trace-Free (PSTF)
tensors by virtue of the recursion relation
 \be\label{recursion}
 \ma{Q}^{\0}_{A_l}(k)=-\f{a}{k}D_{\la
 a_l}\ma{Q}^{\0}_{A_{l-1}\ra}(k)\;.\ee
Another useful relation is
 \be\label{diver}
 D^{a_l}\ma{Q}^{\0}_{A_l}(k)=\f{k}{a}\f{l}{(2l-1)}
 \le[1-(l^2-1)\f{K}{k^2}\ri]\ma{Q}^{\0}_{A_{l-1}}(k)\;,\ee
where the constant $K$ is related to the spatial geometry of the
universe ($K=0,+1,-1$ corresponds to a flat, closed and open
universe, respectively). In the above expressions, we have used
the covariant spherical multipole expansion
 \be
 f(x^a,p^a)=\sum_{l=0}^{\infty}F_{A_l}(x^a,E)e^{A_l}=F(E)+
 F_a(E)e^a+F_{ab}(E)e^ae^b+\cdots\;,\ee
where the PSTF tensor reads
$F_{A_l}(E)=F_{\la a_1a_2\cdots a_l\ra}(E)$.

Next, we expand all dynamical variables in terms of the harmonic
tensors, which is similar to the Fourier series expansion. For the
multipoles of intensity brightness of photon $I_{A_l}$ and
neutrino $G_{A_l}$, we have
 \bea\label{harm_coef_g}
 I_{A_l}&=&I\sum_k\le(\prod_{n=0}^l\kappa_n^{\0}\ri)^{-1}
 \ma{I}_l^{\0}(k)\ma{Q}_{A_l}^{\0}(k)\;,\quad l\geq 1\;,\\
 G_{A_l}&=&G\sum_k\le(\prod_{n=0}^l\kappa_n^{\0}\ri)^{-1}
 \ma{G}_l^{\0}(k)\ma{Q}_{A_l}^{\0}(k)\;,\quad l\geq 1\;.\label{harm_coef_n}\eea
where $I=\r^{(\g)}$, $G=\r^{(\n)}$ and
$\kp_l^{(m)}=\le[1-(l^2-1-m)K/k^2\ri]^{1/2}$ for $l\geq m$. For the
sake of briefness, we will suppress the scalar superscript $(0)$ and
momentum $k$ implicitly in $\ma{Q}_{A_l}^{\0}(k)$,
$\ma{I}_l^{\0}(k)$ and $\ma{G}_l^{\0}(k)$ in the rest part of our
paper. And for other gauge-invariant variables, we have
 \bea
 \D_a^{(i)}&=&\f{aD_a\r^{(i)}}{\r^{(i)}}=-\sum_kk\D^{(i)}_k\ma{Q}_a\;,\label{krho}\\
 q_a^{(i)}&=&\r^{(i)}\sum_kq_k^{(i)}\ma{Q}_a\;,\label{kq}\\
 v_a^{(i)}&=&\sum_kv_k^{(i)}\ma{Q}_a\;,\label{kv}\\
 \pi^{(i)}_{ab}&=&\r^{(i)}\sum_k\pi_k^{(i)}\ma{Q}_{ab}\;,\label{kpi}\\
 \ma{Z}_a&=&-\sum_k\f{k^2}{a}\ma{Z}_k\ma{Q}_{a}\;,\label{kZ}\\
 E_{ab}&=&-\sum_k\f{k^2}{a^2}\ma{E}_k\ma{Q}_{ab}\;,\label{kE}\\
 \s_{ab}&=&-\sum_k\f{k}{a}\s_k\ma{Q}_{ab}\;,\label{ksigma}\\
 A_a&=&\sum_k\f{k}{a}A_k\ma{Q}_a\;,\label{kA}\\
 \pi^{(B)}_{ab}&=&\r^{(\g)}\sum_k\pi^{(B)}_k\ma{Q}_{ab}\;,\label{kpiB}\\
 \D_a^{(B)}&=&\f{aD_a\r^{(B)}}{\r^{(\g)}}=-\sum_kk\D^{(B)}_k\ma{Q}_a\;,\label{krhoB}\eea
where the superscript $(i)$ denotes photon $(\g)$, massless
neutrino $(\n)$, baryon $(b)$ and Cold Dark Matter $(c)$,
respectively. Since $\r^{(B)}$ vanishes at the background level,
in (\ref{krhoB}) and (\ref{kpiB}) we therefore normalize
$\pi^{(B)}_{ab}$ and $\D^{(B)}_a$ by photon density $\r^{(\g)}$.
In the above conventions we have
 \bea
 \ma{I}_0=\D^{(\g)}_k\;,\quad \ma{I}_1=q^{(\g)}_k\;,\quad
 \ma{I}_2=\pi^{(\g)}_k\;,\\
 \ma{G}_0=\D^{(\n)}_k\;,\quad \ma{G}_1=q^{(\n)}_k\;,\quad
 \ma{G}_2=\pi^{(\n)}_k\;.\eea

We are now ready to derive the scalar multipole equations for all
 matter components.
\subsection{Photons}
From (\ref{boltz1}),
the complete Boltzmann hierarchies for the total intensity of photon are
 \bea\label{boltz2}
 &&\dot\ma{I}_l+\f{k}{a}\le[\f{(l+1)}{(2l+1)}\kp_{l+1}^{\0}\ma{I}_{l+1}-\f{l}{(2l+1)}\kp_l^{\0}
 \ma{I}_{l-1}\ri]+4\dot
 h\d_{l0}+\f{4}{3}\f{k}{a}A_k\d_{l1}-\f{8}{15}\f{k}{a}\kp_2^{\0}\s_k\d_{l2}\nn\\
 &&=-n_e\s_T\le[\ma{I}_l-\ma{I}_0\d_{l0}-\f{4}{3}v^{(b)}_k\d_{l1}-\f{1}{10}\ma{I}_2\d_{l2}\ri]\;,\eea
where $\dot h=\le(k\ma{Z}_k/a-\Theta A_k\ri)/3$, the dot is
derivative with respect to cosmic time $t$ and $n_e\s_T$ is the
differential optical depth of the Thompson scattering. The first
three hierarchy equations are \be\label{kgamma Da}
\dot\D^{(\g)}_k+\f{k}{a}\le(\f{4}{3}\ma{Z}_k+q^{(\g)}_k\ri)
 -\f{4}{3}\Theta A_k=0\;,\ee
for  the monopole case ($l=0$),
 \be\label{kgamma l=1} \dot q^{(\g)}_k+\f{1}{3}\f{k}{a}(2\pi^{(\g)}_k-\D^{(\g)}_k+4A_k)
 =n_e\s_T\le(\f{4}{3}v^{(b)}_k-q^{(\g)}_k\ri)\;,\ee
for the diploe case ($l=1)$, and
 \be\label{kgamma l=2}
 \dot\pi^{(\g)}_k+\f{3}{5}\f{k}{a}\ma{I}_3-\f{2}{5}\f{k}{a}q^{(\g)}_k
 -\f{8}{15}\f{k}{a}\s_k=-\f{9}{10}n_e\s_T\pi^{(\g)}_k\;,\ee
for the quadrupole case, respectively.

\subsection{Massless Neutrinos}
Because the massless neutrino only gravitate, the Boltzmann
hierarchies for the total intensity is similar with the one for
photons except that in the right hand side, the Thompson
scattering term vanishes:
 \bea\label{nuboltz2}
 &&\dot\ma{G}_l+\f{k}{a}\le[\f{(l+1)}{(2l+1)}\kp_{l+1}^{\0}\ma{G}_{l+1}-\f{l}{(2l+1)}\kp_l^{\0}
 \ma{G}_{l-1}\ri]\nn\\
 &&+4\dot
 h\d_{l0}+\f{4}{3}\f{k}{a}A_k\d_{l1}-\f{8}{15}\f{k}{a}\kp_2^{\0}\s_k\d_{l2}=0\;.\eea
Because the massless neutrinos behave like collisionless
relativistic particles, we treat them as the improved fluid, i.e.
we need expand in the multipole series to octupole at least. So in
what follows we list the first four hierarchies:

monopole ($l=0$)
 \be\label{knu Da} \dot\D^{(\n)}_k+\f{k}{a}\le(\f{4}{3}\ma{Z}_k+q^{(\n)}_k\ri)
 -\f{4}{3}\Theta A_k=0\;,\ee

dipole ($l=1$)
 \be\label{knu l=1}
 \dot q^{(\n)}_k+\f{1}{3}\f{k}{a}(2\pi^{(\n)}_k-\D^{(\n)}_k+4A_k)=0\;,\ee

quadrupole ($l=2$)
 \be\label{knu l=2}
 \dot\pi^{(\n)}_k+\f{3}{5}\f{k}{a}\ma{G}_3-\f{2}{5}\f{k}{a}q^{(\n)}_k
 -\f{8}{15}\f{k}{a}\s_k=0\;,\ee

octupole ($l=3$)
 \be\label{nu l=3}
 \dot\ma{G}_3=\f{k}{a}\f{3}{7}\pi^{(\n)}_k\;.\ee

\subsection{Bayrons}
For baryons and CDM we use the fluid approximation and neglect their
anisotropic stress tensors, i.e. we characterize baryons and CDM
only by the energy densities and velocities. For baryon density
contrast we have
 \be\label{kbDa}
 \dot\D^{(b)}_k+\le(1+\f{p^{(b)}}{\r^{(b)}}\ri)\le[\f{k}{a}(\ma{Z}_k+v^{(b)}_k)-\T A_k\ri]
 +\le(c_s^2-\f{p^{(b)}}{\r^{(b)}}\ri)\T\D^{(b)}_k=0\;,\ee
where we use $D_ap^{(b)}=c_s^2D_a\r^{(b)}$. The baryon velocity
equation reads
 \bea\label{kbv}
 \le(1+\f{p^{(b)}}{\r^{(b)}}\ri)\le[\dot v^{(b)}_k+\f{1}{3}(1-3c_s^2)\T v^{(b)}_k+\f{k}{a}A_k\ri]
 -\f{k}{a}c_s^2\D^{(b)}_k=\nn\\
 -\f{1}{\r^{(b)}}
 \le[n_e\s_T\r^{(\g)}\le(\f{4}{3}v^{(b)}_k-q^{(\g)}_k\ri)+\f{k\r^{(\g)}}{3a}(2\pi_k^{(B)}-\D^{(B)}_k)\ri]\;,\eea
where the first term in the right hand side denotes the usual
Thompson scattering and the second new term for the Lorentz force
from PMFs.

\subsection{Cold Dark Matter}
For CDM we have
 \be\label{kcDa} \dot\D^{(c)}_k+\f{k}{a}(\ma{Z}_k+v^{(c)}_k)-\T A_k=0\;,\ee
and
 \be\label{kcv} \dot v^{(c)}_k+\f{1}{3}\T
 v^{(c)}_k+\f{k}{a}A_k=0\;,\ee
 respectively.

\subsection{Gravitational equations}
The evolution equations of gravitational field read
 \bea\label{kZ2} &&\dot\ma{Z}_k+\f{\T}{3}\ma{Z}_k+\f{a}{2k}\le[2(\r^{(\g)}\D^{(\g)}_k+\r^{(\n)}\D^{(\n)}_k)+
 \r^{(b)}(1+3c_s^2)\D^{(b)}_k+\r^{(c)}\D^{(c)}_k+2\r^{(\g)}\D^{(B)}_k\ri]\nn\\
 &&-\f{3a}{2k}\le[\f{4}{3}(\r^{(\g)}+\r^{(\n)})+\r^{(c)}+\r^{(b)}+p^{(b)}\ri]A_k-\f{k}{a}A_k=0\;,\eea
 \bea\label{kE2}
 &&\dot\ma{E}_k+\f{\T}{3}\ma{E}_k+\f{a}{2k}\le[(\r+p)\s_k+\r^{(i)}q^{(i)}_k\ri]\nn\\
 &&+\f{a^2}{6k^2}\T\le[3\le(\r^{(i)}+p^{(i)}\ri)-\r^{(i)}\ri]\pi^{(i)}_k
 -\f{a^2}{2k^2}\r^{(i)}\dot\pi^{(i)}_k=0\;,\eea
 \be\label{ksigma2}
 \f{k}{a}\le(\dot\s_k+\f{\T}{3}\s_k\ri)+\f{k^2}{a^2}(\ma{E}_k-A_k)+
 \half\le(\r^{(\g)}\pi^{(\g)}_k+\r^{(\n)}\pi^{(\n)}_k+\r^{(\g)}\pi^{(B)}_k\ri)=0\;.\ee
And the corresponding constraint equations are
 \be\label{constr1} 2\ma{E}_k-\f{a^2}{k^2}\le(\r^{(i)}\pi^{(i)}_k+\r^{(i)}\D^{(i)}_k\ri)-\f{a^3}{k^3}\T
 \r^{(i)}q^{(i)}_k=0\;,\ee
 \be\label{constr2}
 \f{2k^2}{3a^2}(\ma{Z}_k-\s_k)+\r^{(i)}q^{(i)}_k=0\;.\ee

\section{\label{ic}Compensated magnetic initial conditions and CMB power spectrum}
In this section, we analytically extract the scalar modes seeded
by PMFs in the deep radiation dominated era. Then we numerically
calculate the CMB angular power spectrum by using two compensated
magnetic initial conditions.

\subsection{Equations in the tight-coupling approximation}
In this subsection we propagate the covariant equations in the zero-acceleration
frame, in which the CDM velocity vanishes.
For the density contrast of different species, we have
 \be\label{ic_den_g}
 \D^{(\g)'}_k+k\le(\f{4}{3}\ma{Z}_k+q^{(\g)}_k\ri)=0\;,\ee
 \be\label{ic_den_n}
 \D^{(\n)'}_k+k\le(\f{4}{3}\ma{Z}_k+q^{(\n)}_k\ri)=0\;,\ee
 \be\label{ic_den_b}
 \D^{(b)'}_k+k\le(\ma{Z}_k+v^{(b)}_k\ri)=0\;,\ee
 \be\label{ic_den_c}
 \D^{(c)'}_k+k\ma{Z}_k=0\;,\ee
where $'=d/d\eta$ is the derivative with respect to the conformal
time $\h$. For simplicity, we also set the pressure and sound-speed
of baryon fluid to zero ($p^{(b)}=c_s^2=0$).

In the deep radiation dominant era,  photons are tightly coupled
with ionized baryons through the Thomspon scattering. This allows
us to deal with them as a single baryon-photon fluid with a common
fluid velocity $q^{(\g b)}_k$. Furthermore, PMFs also exert the
Lorentz force onto the baryon-photon fluid. Hence, under the
tight-coupling approximation ($q^{(\g b)}_k\simeq q^{(\g)}_k\simeq
4v^{(b)}_k/3$), the velocity equation of the baryon-photon fluid
takes the following form
 \be\label{ic_q_g}
 q^{(\g b)'}_k+\f{\ma{H}}{(1+R)}q^{(\g b)}_k-\f{kR}{3(1+R)}\D^{(\g)}_k
 +\f{kR}{3(1+R)}(2\pi^{(B)}_k-\D^{(B)}_k)=0\;,\ee
where $\ma{H}=a'/a$ is the conformal Hubble constant and
$R=4\r^{(\g)}/3\r^{(b)}$ is the photon to baryon ratio.

Since the massless neutrinos behave as collisionless relativistic
particles, they can preserve the non-vanishing octupole signals
 \be\label{ic_q_n}
 q^{(\n)'}_k+\f{k}{3}(2\pi^{(\n)}_k-\D^{(\n)}_k)=0\;,\ee
 \be\label{ic_pi_n}
 \pi^{(\n)'}_k+\f{3}{5}k\ma{G}_3-\f{2}{5}kq^{(\n)}_k-\f{8}{15}k\s_k=0\;,\ee
 \be\label{ic_G_3}
 \ma{G}'_3=\f{3}{7}k\pi^{(\n)}_k\;.\ee

The gravitational field equations in the zero-acceleration frame
read
 \be\label{ic_Z}
 \ma{Z}'_k+\ma{H}\ma{Z}_k+\f{3\ma{H}^2}{k}\le[R_{\g}\D^{(\g)}_k+R_{\n}\D^{(\n)}_k
 +R_{\g}\D^{(B)}_k+\half R_c\D^{(c)}_k+\half R_b\D^{(b)}_k\ri]=0\;,\ee
 \bea\label{ic_weyl}
 &&\ma{E}'_k+\ma{H}\ma{E}_k+\f{3\ma{H}^2}{2k}\le[\f{4}{3}\s_k+R_{\g}q^{(\g)}_k+R_{\n}q^{(\n)}_k+R_bv^{(b)}_k\ri]
 +\f{9\ma{H}^3}{2k^2}\le[R_{\g}\pi^{(B)}_k+R_{\n}\pi^{(\n)}_k\ri]\nn\\
 &&-\f{3\ma{H}^2}{2k^2}\le[R_{\g}\pi^{(B)'}_k+R_{\n}\pi^{(\n)'}_k\ri]=0\;,\eea
 \be\label{ic_sigma}
 \s'_k+\ma{H}\s_k+k\ma{E}_k+\f{3\ma{H}^2}{2k}\le[R_{\g}\pi^{(B)}_k+R_{\n}\pi^{(\n)}_k\ri]=0\;,\ee
where we define the density fraction as $R_{\g}=\r_{\g}/\r$,
$R_{\n}=\r_{\n}/\r$, $R_{b}\h=\r_{b}/\r$ and $R_{c}\h=\r_{c}/\r$.
Note that in our definitions $R_b$ and $R_c$ have the dimension
$({\rm length})^{-1}$. In addition, the gravitational constraint
equations are
 \bea\label{ic_con1}
 &&2\ma{E}_k-\f{3\ma{H}^2}{k^2}\le[R_{\g}\pi^{(B)}_k+R_{\n}\pi^{(\n)}_k+R_{\g}\D^{(\g)}_k
 +R_{\g}\D^{(B)}_k+R_{\n}\D^{(\n)}_k+
 R_c\D^{(c)}_k+R_b\D^{(b)}_k\ri]\nn\\
&&-\f{9\ma{H}^3}{k^3}\le[R_{\g}q^{(\g)}_k+R_{\n}q^{(\n)}_k+R_bv^{(b)}_k\ri]
 =0\;,\eea
 \be\label{ic_con2}
 \ma{Z}_k-\s_k+\f{9\ma{H}^2}{2k^2}\le[R_{\g}q^{(\g)}_k+R_{\n}q^{(\n)}_k+R_bv^{(b)}_k\ri]=0\;.\ee

\subsection{Compensated magnetic initial conditions}
Since the radiation species dominate our universe during the
initial era ($R_{\g}+R_{\n}\simeq 1$), usually one neglects the
matter contributions when derives the adiabatic initial
conditions. However, as demonstrated in
\cite{Kojima:2009ms,Shaw:2009nf}, one cannot neglect the matter
contributions in the case of existence of PMFs due to the
compensation mechanism between the radiation density perturbations
and those of PMFs. So it is essential to take the matter
contributions into account when we derive the magnetic initial
conditions. In addition, it turns out convenient to introduce a
new characteristic length scale $R_m=R_b+R_c\simeq
\r_m(\h_0)/\sqrt{3\r_r(\h_0)}\simeq 5\times10^{-3} {\rm
Mpc}^{-1}$. In what follows, we list two different compensated
magnetic modes including matter contributions.

The density $\D^{(B)}_k$ sourced compensated magnetic mode:
 \bea\label{ic_density}
 \D^{(\g)}_k&=&-R_{\g}+\f{R_{\g}R_m}{2k}k\h
 -\le[\f{R_{\n}}{6}+\f{3R_{\g}R_m^2}{16k^2}\ri]k^2\h^2\;,\\
 \D^{(\n)}_k&=&-R_{\g}+\f{R_{\g}R_m}{2k}k\h
 +\le[\f{R_{\g}}{6}-\f{3R_{\g}R_m^2}{16k^2}\ri]k^2\h^2\;,\\
 \D^{(b)}_k&=&-\f{3}{4}R_{\g}+\f{3R_{\g}R_m}{8k}k\h
 -\le[\f{R_{\n}}{8}+\f{9R_{\g}R_m^2}{64k^2}\ri]k^2\h^2\;,\\
 \D^{(c)}_k&=&-\f{3}{4}R_{\g}+\f{3R_{\g}R_m}{8k}k\h
 -\f{9R_{\g}R_m^2}{64k^2}k^2\h^2\;,\\
 q^{(\g)}_k&=&\f{R_{\n}}{3}k\h+\le[\f{R_mR_{\g}}{12k}
 -\f{R_bR_{\n}}{4kR_{\g}}\ri]k^2\h^2\;,\\
 q^{(\n)}_k&=&-\f{R_{\g}}{3}k\h+\f{R_{\g}R_m}{12k}k^2\h^2\;,\\
 \pi^{(\n)}_k&=&-\f{R_{\g}}{15+4R_{\n}}k^2\h^2\;,\\
 \ma{G}_3&=&-\f{3R_{\g}}{7(15+4R_{\n})}k^3\h^3\;,\\
 \h_s&=&\f{R_{\g}R_m}{8k}k\h+\le[\f{R_{\n}R_{\g}}{6(15+4R_{\n})}
 -\f{3R_{\g}R_m^2}{64k^2}\ri]k^2\h^2\;.
 \eea
And the anisotropic stress tensor $\pi^{(B)}_k$ sourced mode:
 \bea\label{ic_piB1}
 \D^{(\g)}_k&=&\f{1}{3}k^2\h^2\;,\\
 \D^{(\n)}_k&=&-\f{R_{\g}}{3R_{\n}}k^2\h^2\;,\label{ic_piB2}\\
 \D^{(b)}_k&=&\f{1}{4}k^2\h^2\;,\label{ic_piB3}\\
 \D^{(c)}_k&=&-\f{R_b}{40k}k^3\h^3\;,\label{ic_piB4}\\
 q^{(\g)}_k&=&-\f{2}{3}k\h+\f{R_b}{2R_{\g}k}k^2\h^2\;,\label{ic_piB5}\\
 q^{(\n)}_k&=&\f{2R_{\g}}{3R_{\n}}k\h\;,\label{ic_piB6}\\
 \pi^{(\n)}_k&=&-\f{R_{\g}}{R_{\n}}+\f{55R_{\g}}{14R_{\n}(15+4R_{\n})}k^2\h^2\;,\label{ic_piB7}\\
 \ma{G}_3&=&-\f{3R_{\g}}{7R_{\n}}k\h\;,\label{ic_piB8}\\
 \h_s&=&-\f{55R_{\n}}{84(15+4R_{\n})}k^2\h^2\;,\label{ic_piB9}
 \eea
where $\h_s=-(2\ma{E}_k+\s'_k/k)$ is the curvature perturbation in
the synchronous gauge.

\subsection{\label{cmb_pow_spec}CMB power spectrum}
By virtue of the above initial conditions, we formally integrate the
set of evolution equations over the line of sight
\cite{Seljak:1996is,Hu:1997mn,Hu:1997hp,Challinor:2000as,Zaldarriaga:1997va}
 \bea\label{intboltz}
 \ma{I}_l&=&4\int^{t_R}dt e^{-\tau}\le\{\le[\f{k}{a}\s_k+\f{3}{16}n_e\s_T(\kp_2^{\0})^{-1}
 \ma{I}_2\ri]\le[\f{1}{3}j_l(x)+\f{\dd^2}{\dd x^2}j_l(x)\ri]\ri.\nn\\
 &&\le.-\le(\f{k}{a}A_k-n_e\s_Tv_k\ri)\f{\dd}{\dd x}j_l(x)-\le[\f{1}{3}\le(\f{k}{a}\ma{Z}_k-\Theta A_k\ri)
 -\f{1}{4}n_e\s_T\ma{I}_l\ri]
 j_l(x)\ri\}\;,\eea
where $\tau=\int n_e\s_T\dd t$ is the optical depth, $x=k\chi$ with
$\chi$ is the comoving radial distance along the line of sight and
$j_{l}(x)$ are the spherical Bessel functions.

 \begin{figure}[h]
    \centering
    \includegraphics[angle=-90,width=11.5cm]{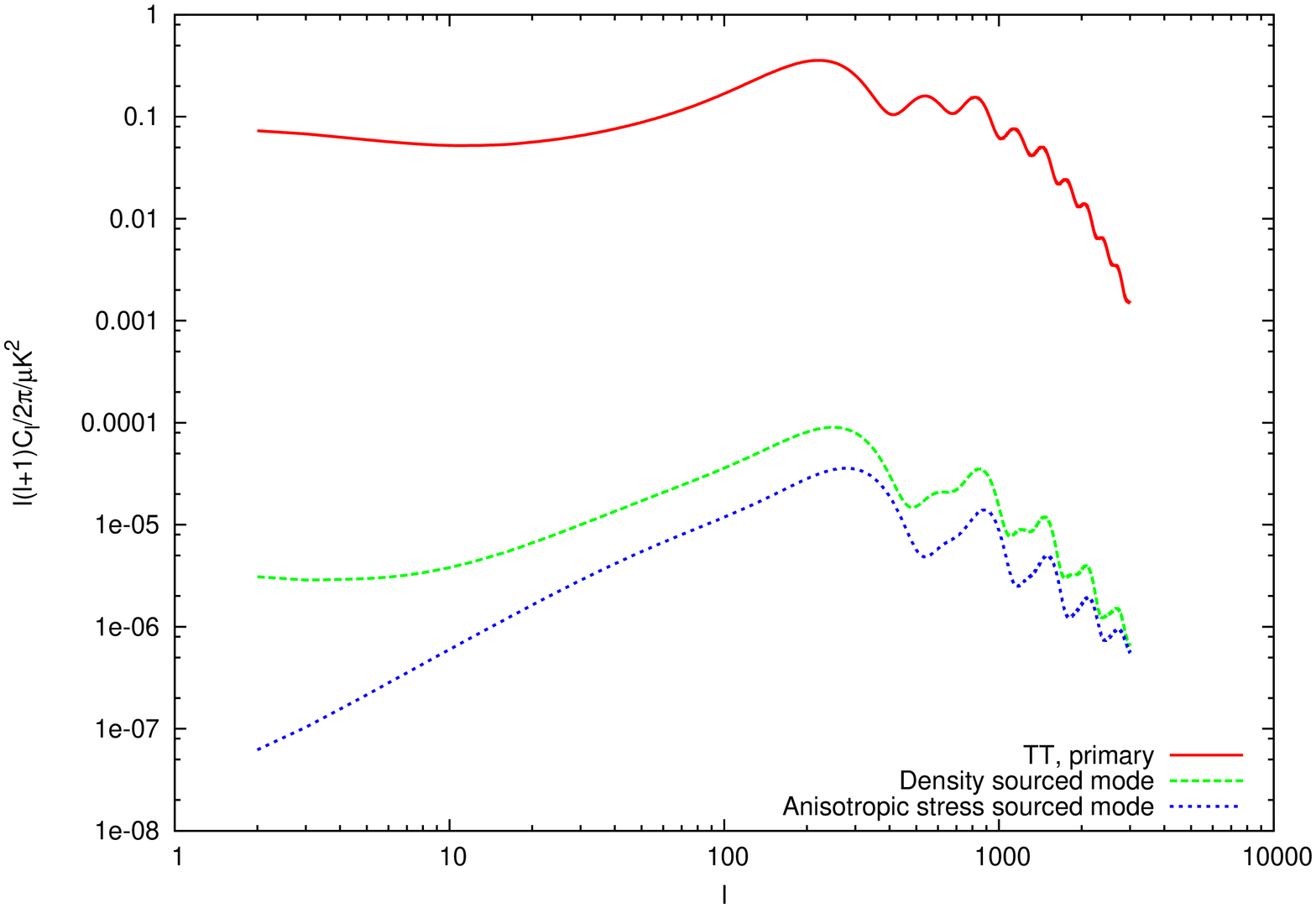}
    \caption{The CMB spectrum of TT mode with the magnetic index $n_B=-2.9$ and
 comoving magnetic mean-field amplitude $B_{\lam}=9{\rm~nG}$. The red solid curve
 stands for the primary adiabatic mode, the green dashed one for the magnetic
density $\D^{(B)}_k$
 sourced mode and the blue dotted one for the magnetic anisotropic stress $\pi^{(B)}_k$
 sourced mode, respectively.}
    \label{tt}
 \end{figure}

 \begin{figure}[h]
    \centering
    \includegraphics[angle=-90,width=11.5cm]{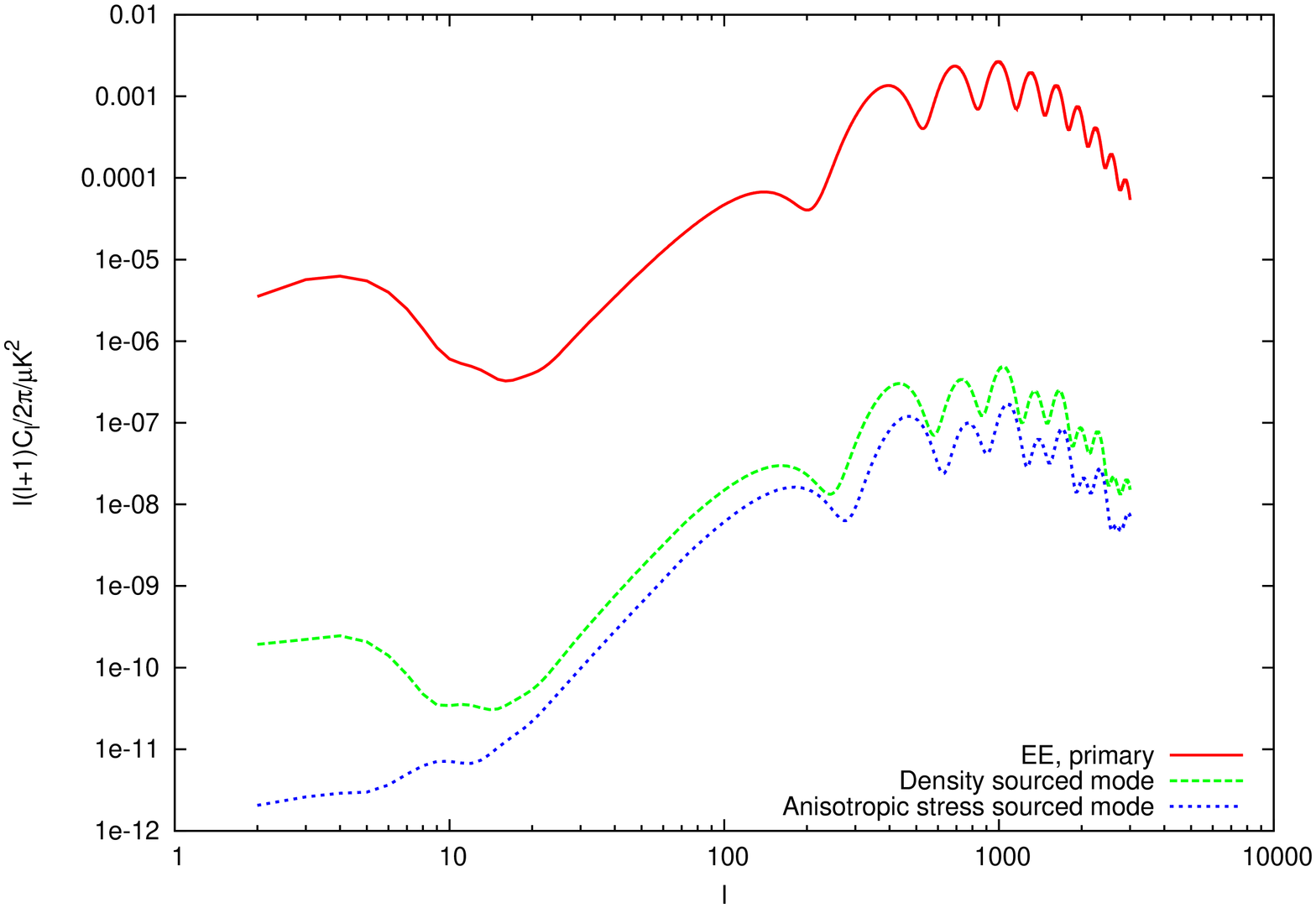}
    \caption{The CMB spectrum of EE mode with  magnetic index $n_B=-2.9$ and
 comoving magnetic mean-field amplitude $B_{\lam}=9{\rm~nG}$. The red solid curve
 stands
 for the primary adiabatic mode, the green dashed one for the magnetic density $\D^{(B)}_k$
 sourced mode and the blue dotted one for the magnetic anisotropic stress $\pi^{(B)}_k$
 sourced mode, respectively.}
    \label{ee}
 \end{figure}

 \begin{figure}[h]
    \centering
    \includegraphics[angle=-90,width=11.5cm]{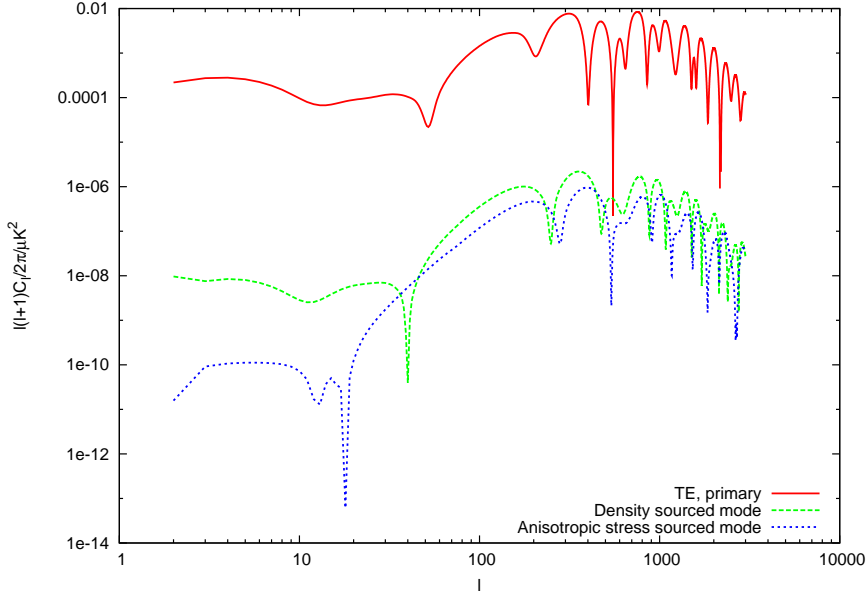}
    \caption{The CMB spectrum of TE mode with magnetic index $n_B=-2.9$ and
 comoving magnetic mean-field amplitude $B_{\lam}=9{\rm~nG}$. The red solid curve
 stands
 for the primary adiabatic mode, the green dashed one for the magnetic density $\D^{(B)}_k$
 sourced mode and the blue dotted one for the magnetic anisotropic stress $\pi^{(B)}_k$
 sourced mode, respectively.}
    \label{te}
 \end{figure}

Then we expand the temperature contrast ($\d_{T}=\d T/T_0$) in the
multipole series
 \be\label{dToverT1} \d_T(e^a)
 =\f{\pi}{I}\sum_{l=1}^{\infty}\D_l^{-1}I_{A_l}e^{A_l}
 =\pi\sum_{l=1}^{\infty}\sum_k\D_l^{-1}
 C_kg_{Tl}(k)\ma{Q}_{A_l}e^{A_l}\;,\ee
where in the second equality we rewrite the multipole coefficient
$\ma{I}_l(k)=C_kg_{Tl}(k)$ with the transfer function $g_{Tl}(k)$
and random variables $C_k$, which source the CMB anisotropies with
primordial power spectrum
 \be\label{prim_pow}
 \la C_kC_{k'}^{\ast}\ra=C^2(k)\d_{kk'}\;.\ee
Usually $C_k$ are the primordial curvature perturbations, however,
in this paper they are the density contrast $\D^{(B)}_k$ or
anisotropic stress $\pi^{(B)}_k$ of PMFs, and their power
spectra are given in (\ref{pow_DB}) and (\ref{pow_piB}).

Armed with the primordial power spectrum, we finally obtain the CMB
angular power spectrum
 \be\label{Cl}
 C_l=\pi^2\int_0^{\infty} d\ln k~C^2(k)|g_{Tl}(k)|^2\;.\ee
In Figure \ref{tt}, \ref{ee} and \ref{te}, we plot the CMB TT, EE
and TE spectra with the primary adiabatic mode (red solid curve),
magnetic density $\D^{(B)}_k$ sourced mode (green dashed one) and
magnetic anisotropic stress $\pi^{(B)}_k$ sourced mode (blue
dotted one), respectively. In our numerical calculations, we
modify CAMB code \cite{Lewis:1999bs} and set the amplitudes of
primordial curvature perturbations to unit.  The results show that
PMFs contribute a tiny part to the CMB power spectra, however, in
the next section,  we will demonstrate that they will give a
dominant contribution at the bispectrum level.

\section{\label{bispectrum}CMB bispectrum}
In this section we  numerically calculate the CMB bispectrum
 seeded by the compensated magnetic density mode.

\subsection{Analytic formulas}
Firstly, let us shortly review the analytic formulas to calculate
CMB angular bispectrum
\cite{Wang:1999vf,Hu:2000ee,Komatsu:2001rj}. In (\ref{dToverT1})
we decompose the temperature contrast $\d_T$ by the covariant
approach, now we expand it in terms of the spherical harmonics,
which are more familiar to us,
 \be\label{dToverT2}
 \d_T(\hat\bn)=\sum_{lm}a_{lm}Y_{lm}(\hat\bn)\;,\ee
where $\hat\bn$ denotes the unit direction vector. The CMB angular
bispectrum is defined as
 \be\label{bi1}
 B^{m_1m_2m_3}_{l_1l_2l_3}\equiv\la a_{l_1m_1}a_{l_2m_2}a_{l_3m_3}\ra\;,\ee
where $B^{m_1m_2m_3}_{l_1l_2l_3}$ must satisfy the triangle
conditions and the selection rules: $m_1+m_2+m_3=0$,
$l_1+l_2+l_3={\rm~even}$ and $|l_i-l_j|\leq l_k\leq l_i+l_j$ for
all permutations of indices. Note that Gaunt integral satisfies
all the conditions mentioned above
 \bea\label{Gaunt}
 \ma{G}^{m_1m_2m_3}_{l_1l_2l_3}&\equiv&\int d^2\hat\bn~Y_{l_1m_1}(\hat\bn)
 Y_{l_2m_2}(\hat\bn)Y_{l_3m_3}(\hat\bn)\;,\nn\\
 &=&\sqrt{\f{(2l_1+1)(2l_2+1)(2l_3+1)}{4\pi}}
 \le(
 \begin{array}{ccc}
  l_1&l_2&l_3\\
  0&0&0\\
 \end{array}
 \ri)\times
 \le(
 \begin{array}{ccc}
  l_1&l_2&l_3\\
  m_1&m_2&m_3\\
 \end{array}
 \ri)\;,
 \eea
 where the matrices denote the Wigner-3j symbol.
Therefore it is convenient to introduce the reduced bispectrum
(Komatsu-Spergel estimator) $b_{l_1l_2l_3}$ \cite{Komatsu:2001rj}
to replace $B^{m_1m_2m_3}_{l_1l_2l_3}$ without any loss of
information
 \be\label{red_bi}
 B^{m_1m_2m_3}_{l_1l_2l_3}=\ma{G}^{m_1m_2m_3}_{l_1l_2l_3}b_{l_1l_2l_3}\;.\ee
Thus, the observable angle-averaged bispectrum can be written as
 \bea
 B_{l_1l_2l_3}&\equiv&\sum_{m_1m_2m_3}
 \le(
 \begin{array}{ccc}
  l_1&l_2&l_3\\
  m_1&m_2&m_3\\
 \end{array}
 \ri)
 B^{m_1m_2m_3}_{l_1l_2l_3}\nn\\
 &=&\sqrt{\f{(2l_1+1)(2l_2+1)(2l_3+1)}{4\pi}}
  \le(
 \begin{array}{ccc}
  l_1&l_2&l_3\\
  0&0&0\\
 \end{array}
 \ri)
 b_{l_1l_2l_3}\;.\eea

In order to calculate the reduced bispectrum $b_{l_1l_2l_3}$, we
need obtain the form of primordial bispectrum $F(k_1,k_2,k_3)$.
For a slow roll inflation model, the non-gaussian curvature
perturbations $\zeta(\bx)$ are usually parameterized by a single
constant parameter $f^{{\rm local}}_{NL}$ and the Gaussian random
variable $\zeta_L(\bx)$ in the real space
 \be
 \zeta(\bx)=\zeta_L(\bx)+
 f^{{\rm local}}_{NL}\Big[\zeta^2_L(\bx)-\la\zeta^2(\bx)\ra\Big]\;.\ee
In the Fourier space, the local-type primordial curvature
bispectrum $\la \z(\bk_1)\z(\bk_2)\z(\bk_3)\ra\equiv
F_{\zeta}(k_1,k_2,k_3)\d(\bk_1+\bk_2+\bk_3)$ can be obtained by
performing the momentum convolution
 \be\label{F_z}
 F_{\zeta}(k_1,k_2,k_3)\propto f^{{\rm local}}_{NL}
 \Big\{P_{\zeta}(k_1)P_{\zeta}(k_2)k_3^3+(k_1,k_2,k_3){\rm ~perm.}\Big\}\;,\ee
where $P_{\zeta}$ is the primordial power spectrum of curvature
perturbations ($\zeta$). However, we are interested in the PMF
signals in CMB bispectrum which have the essential non-Gaussian
characters. The magnetic density contrast bispectrum $\la
\D^{(B)}(\bk_1)\D^{(B)}(\bk_2)\D^{(B)}(\bk_3)\ra
\equiv F_{\D_B}(k_1,k_2,k_3)\d(\bk_1+\bk_2+\bk_3)$ has been derived
analytically in \cite{Caprini:2009vk,Seshadri:2009sy}
 \bea
 F_{\D_B}(k,q,p)&=&\f{3A^3}{48\pi^2\r^{(\g)^3}}\le[f^{(1)}(k,q,p)
 +f^{(2)}(k,q,p)+f^{(3)}(k,q,p)\ri]\;,\label{F}\\
 f^{(1)}(k,q,p)&=&\f{n_B}{(n_B+3)(2n_B+3)}k^{2n_B+6}q^{n_B+3}p^3
 +(k,q,p){\rm ~perm.}\;,\label{f1}\\
 f^{(2)}(k,q,p)&=&\f{n_B}{(3n_B+3)(2n_B+3)}k^{3}q^{3n_B+6}p^3
+(k,q,p){\rm ~perm.}\;,\label{f2}\\
 f^{(3)}(k,q,p)&=&\f{k_D^{3n_B+3}}{3n_B+3}k^3q^3p^3
 +(k,q,p){\rm ~perm.}\;.\label{f3}\eea
From the above expressions, we notice that the first magnetic
shape $f^{(1)}(k,q,p)$ looks similar to the one from local-type
primordial curvature perturbations (\ref{F_z}), if the magnetic
index takes the nearly scale-invariant value $n_B\simeq-3$. So, we
will abuse the phrase ``local-type'' for the magnetic shape
$f^{(1)}(k,q,p)$ in this paper.

Now, we are ready to calculate the CMB bispectrum sourced by
the magnetic density contrast $\D^{(B)}(\bk)$. Following the
standard procedure \cite{Wang:1999vf}, the reduced bispectrum
can be expressed as
 \bea\label{red_bi2}
 b_{l_1l_2l_3}&=&
 (8\pi)^3\int_0^{\infty}x^2dx\int_0^{k_D}d\ln k\int_0^{k_D}d\ln q
 \int_0^{k_D}d\ln p~j_{l_1}(kx)j_{l_2}(qx)j_{l_3}(px)\nn\\
 && \times F_{\D_B}(k,q,p)g_{Tl_1}(k)g_{Tl_2}(q)g_{Tl_3}(p)\;,
 \eea
where $g_{Tl}(k)$ is the transfer function and $x$ is the comoving
radial distance along the line of sight. From (\ref{f1})-(\ref{f3}),
we can see that the integral (\ref{red_bi2}) is determined by four
kinds of momentum integrations
 \bea
 b^{(\a)}_l(x)&\equiv&\int_0^{k_D}d\ln k~k^{2n_B+6}j_l(kx)g_{Tl}(k)\;,\label{b_la}\\
 b^{(\b)}_l(x)&\equiv&\int_0^{k_D}d\ln k~k^{n_B+3}j_l(kx)g_{Tl}(k)\;,\label{b_lb}\\
 b^{(\g)}_l(x)&\equiv&\int_0^{k_D}d\ln k~k^{3}j_l(kx)g_{Tl}(k)\;,\label{b_lc}\\
 b^{(\d)}_l(x)&\equiv&\int_0^{k_D}d\ln k~k^{3n_B+6}j_l(kx)g_{Tl}(k)\;.\label{b_ld}\eea
Then, we can express the reduced bispectrum (\ref{red_bi2}) in the
following form
 \bea\label{blll}
 b_{l_1l_2l_3}&=&b^{(1)}_{l_1l_2l_3}+b^{(2)}_{l_1l_2l_3}+b^{(3)}_{l_1l_2l_3}\;,\\
 b^{(1)}_{l_1l_2l_3}&=&\int_0^{\infty}x^2dx~\ma{N}_1
 \Big\{b^{(\a)}_{l_1}(x)b^{(\b)}_{l_2}(x)b^{(\g)}_{l_3}(x)
 +(l_1,l_2,l_3){\rm ~perm.}\Big\}\;,\label{blll1}\\
 \ma{N}_1&=&\f{3(8\pi)^3A^3n_B}{48\pi^2(n_B+3)(2n_B+3)\r^{(\g)^3}}\;,\nn\\
 b^{(2)}_{l_1l_2l_3}&=&\int_0^{\infty}x^2dx~\ma{N}_2
 \Big\{b^{(\g)}_{l_1}(x)b^{(\d)}_{l_2}(x)b^{(\g)}_{l_3}(x)
 +(l_1,l_2,l_3){\rm ~perm.}\Big\}\;,\label{blll2}\\
 \ma{N}_2&=&\f{3(8\pi)^3A^3n_B}{48\pi^2(3n_B+3)(2n_B+3)\r^{(\g)^3}}\;,\nn\\
 b^{(3)}_{l_1l_2l_3}&=&\int_0^{\infty}x^2dx~\ma{N}_3
 \Big\{b^{(\g)}_{l_1}(x)b^{(\g)}_{l_2}(x)b^{(\g)}_{l_3}(x)
 +(l_1,l_2,l_3){\rm ~perm.}\Big\}\;,\label{blll3}\\
 \ma{N}_3&=&\f{3(8\pi)^3A^3k_D^{3n_B+3}}{48\pi^2(3n_B+3)\r^{(\g)^3}}\;.\eea
The seven-year WMAP data \cite{Komatsu:2010fb} give
$P_{\z}\sim2.441\times 10^{-9}$.  Hence, from (\ref{F_z}) we can
estimate the amplitude of the primordial curvature bispectrum
$F_{\zeta}(k_1,k_2,k_3)$ is of the order $\ma{O}(10^{-18})$. On
the other hand, the numerical calculations about the magnetic
power spectrum show that the amplitude of $P_{\D_B}$ is
approximately of the order $\ma{O}(10^{-13})$, i.e.
$\D^{(B)}_k\sim \ma{O}(10^{-6})$. Since the some quantities
related to the PMFs such as $\D^{(B)}_k$ and $\pi^{(B)}_k$ have
the essential non-Gaussian characters, the primordial magnetic
bispectrum $F_{\D_B/\pi_B}(k_1,k_2,k_3)$ is proportional to the
cubic of $\D^{(B)}_k$ and $\pi^{(B)}_k$. So, for the magnetic
density sourced mode, the amplitude of bispectrum
$F_{\D_B}(k_1,k_2,k_3)\sim \ma{O}(10^{-18})$ is comparable with
that of the primordial curvature one $F_{\zeta}(k_1,k_2,k_3)\sim
\ma{O}(10^{-18})$. In fact, this observation is just our
motivation for this paper.

 \begin{figure}[h]
    \centering
    \includegraphics[angle=-90,width=11.5cm]{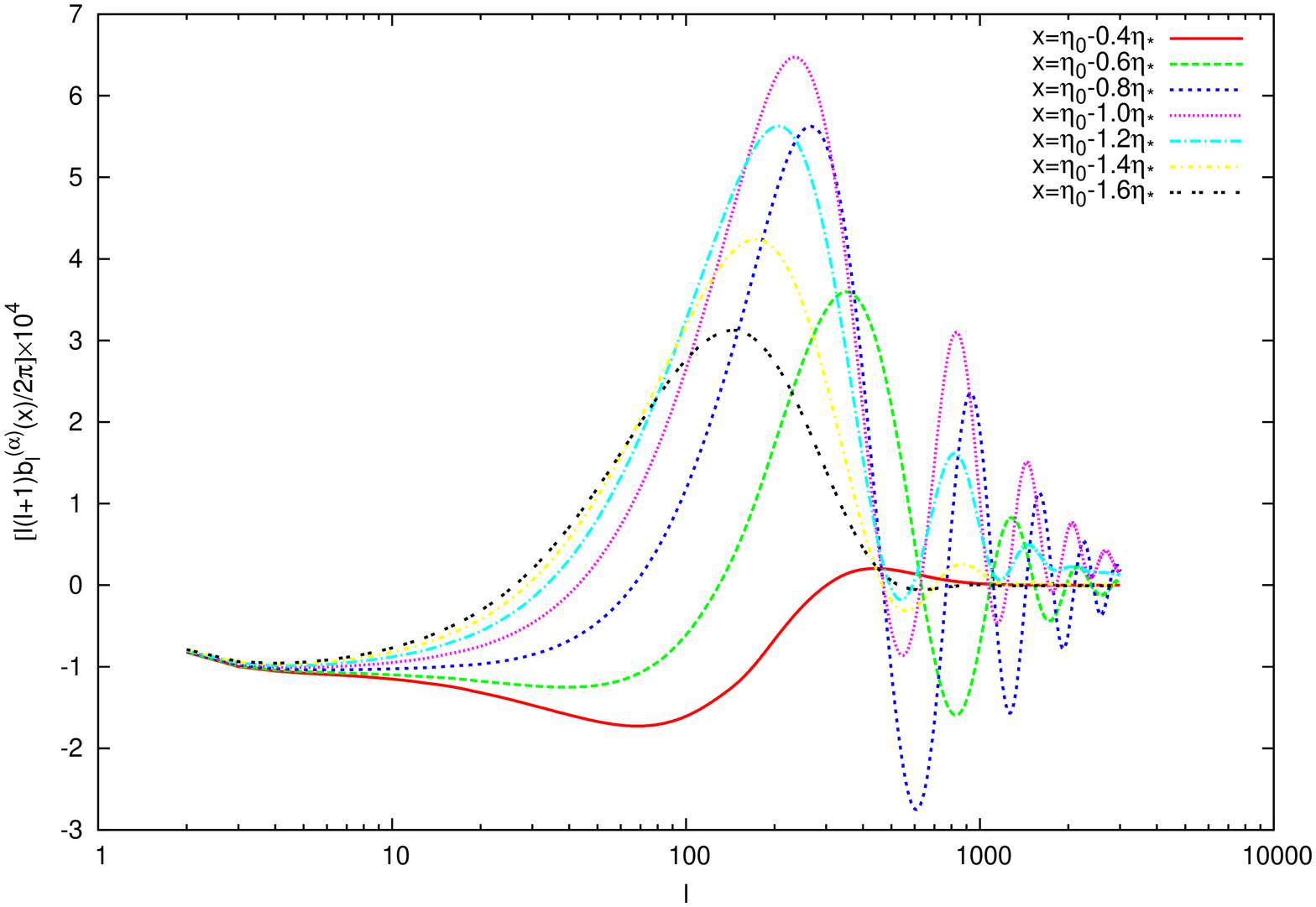}
    \caption{This figure shows $\Big[l(l+1)b^{(\a)}_l(x)/2\pi\Big]\times 10^{4}$
for several different comoving radial distances
$x=(\h_0-0.4\h_{\ast})\sim (\h_0-1.6\h_{\ast})$, where we set
the conformal time at present $\h_0=14.38{\rm~Gpc}$ and at the
recombination epoch $\h_{\ast}=284.85{\rm~Mpc}$, respectively.
Parameters for PMFs are $n_B=-2.9$ and $B_{\lam}=9{\rm~nG}$.}
    \label{ba}
 \end{figure}

 \begin{figure}[h]
    \centering
    \includegraphics[angle=-90,width=11.5cm]{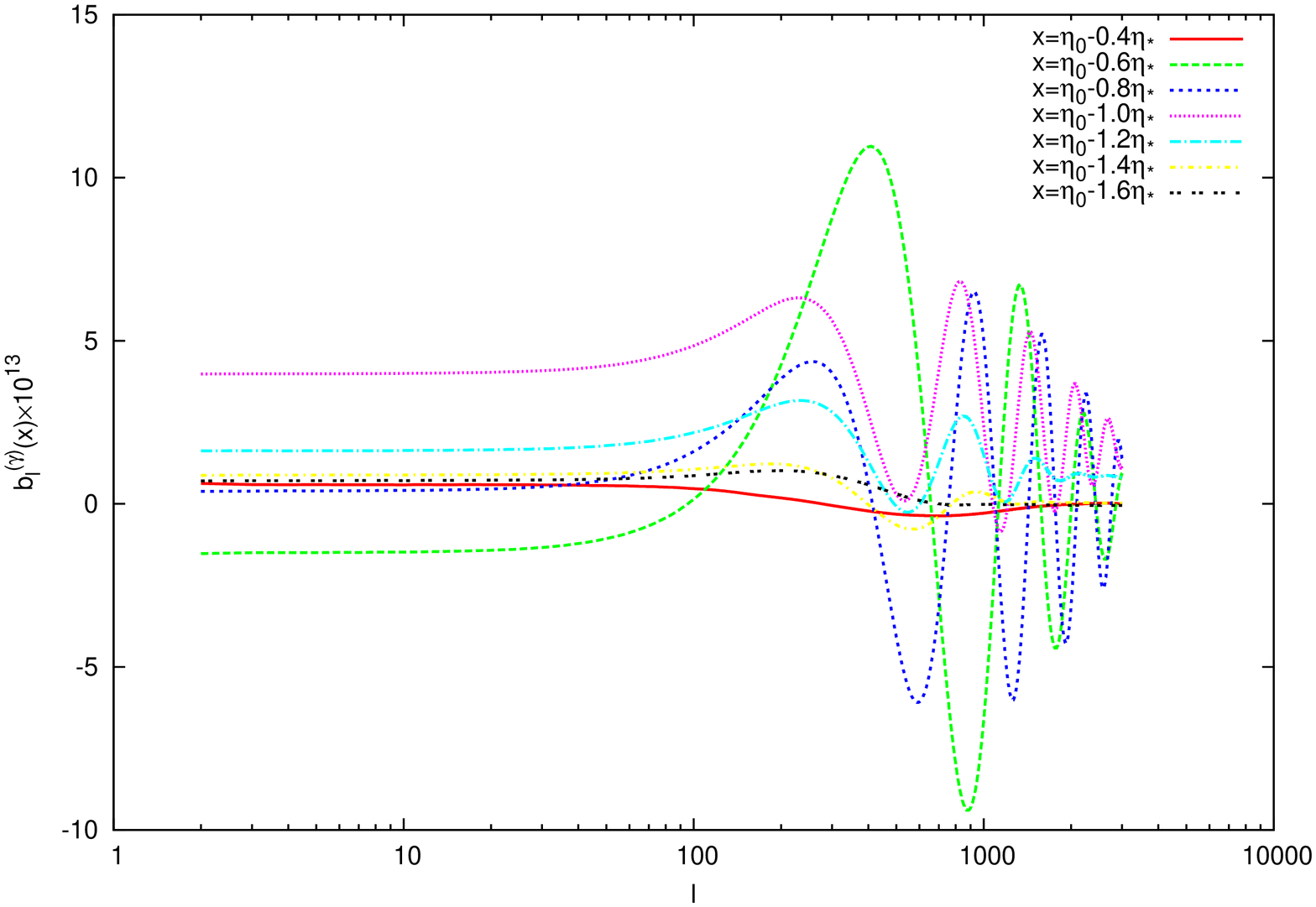}
    \caption{This figure shows $b^{(\g)}_l(x)\times 10^{13}$
for several different comoving radial distances
$x=(\h_0-0.4\h_{\ast})\sim (\h_0-1.6\h_{\ast})$, where we set
the conformal time at present $\h_0=14.38{\rm~Gpc}$ and at the
recombination epoch $\h_{\ast}=284.85{\rm~Mpc}$, respectively.
Parameters for PMFs are $n_B=-2.9$ and $B_{\lam}=9{\rm~nG}$.}
    \label{bc}
 \end{figure}

 \begin{figure}[h]
    \centering
    \includegraphics[angle=-90,width=11.5cm]{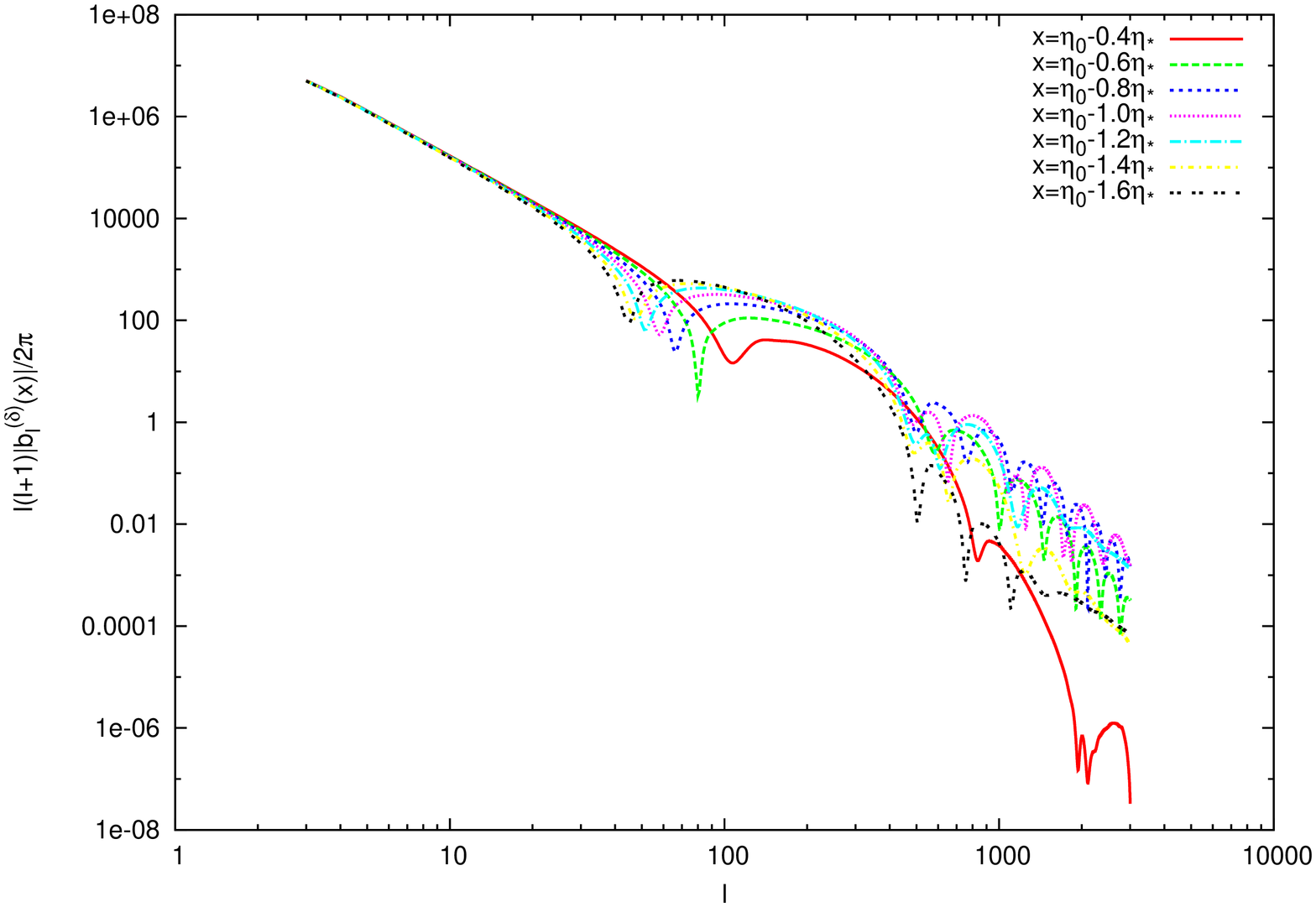}
    \caption{This figure shows $l(l+1)|b^{(\d)}_l(x)|/2\pi$
for several different comoving radial distances
$x=(\h_0-0.4\h_{\ast})\sim (\h_0-1.6\h_{\ast})$, where we set
the conformal time at present $\h_0=14.38{\rm~Gpc}$ and at the
recombination epoch $\h_{\ast}=284.85{\rm~Mpc}$, respectively.
Parameters for PMFs are $n_B=-2.9$ and $B_{\lam}=9{\rm~nG}$.}
    \label{bd}
 \end{figure}

\subsection{Numerical results}
In this subsection, we will present our numerical results about
the reduced bispectra. For the case with $n_B=-2.9$ and
$B_{\lam}=9{\rm~nG}$, the momentum integral (\ref{b_la}) is
approximately equivalent to (\ref{b_lb}), while (\ref{b_ld})
diverges in the infrared limit ($k\rightarrow0$). In Figure
(\ref{ba}), (\ref{bc}) and (\ref{bd}), we plot $b^{(\a)}_l(\simeq
b^{(\b)}_l)$, $b^{(\g)}_l$ and the absolute value of $b^{(\d)}_l$
over $l$ with $\h_0=14.38{\rm~Gpc}$ and
$\h_{\ast}=284.85{\rm~Mpc}$ being the conformal time at present
and at the recombination epoch, respectively. For $n_B=-2.9$, the
momentum shapes in $b^{(\a/\b)}_l$ are nearly scale invariant, so
the profile of $b^{(\a)}_l$-curve (Figure \ref{ba}) looks similar
to those of $C_l$. However, the main difference between
$b^{(\a)}_l$ and $C_l$ is that the former changes the sign, while
the latter does not. The reason lies in that $b^{(\a)}_l\propto
j_l(kx)g_{Tl}(k)$, but $C_l\propto |g_{Tl}(k)|^2$. As what happens
to the standard model
\cite{Komatsu:2001rj,Babich:2004yc,Komatsu:2002db,Fergusson:2006pr},
from Figure (\ref{bc}) we can see that the phase of $b^{(\g)}_l$
in the high-$l$ regime oscillates rapidly with respect to $x$,
which will heavily suppress the integrations (\ref{blll1}),
(\ref{blll2}) and (\ref{blll3}) at  small scales. However, for the
case with the magnetic index $n_B\simeq-3$, the integration
(\ref{b_ld}) is dangerous. Because the exponent of the shape
$k^{3n_B+6}$ is much less than zero, the full integrand will
diverge in the Infrared (IR) limit, if the spherical Bessel
function $j_l(kx)$ and transfer function $g_{Tl}(k)$ cannot
provide an enough positive power to compensate the negative slope
of $k^{3n_B+6}$.  Our numerical calculation indeed shows that such
divergence does exist. Figure (\ref{bd}) illustrates the IR
divergence of $b^{(\d)}_l$, which will dominate over all the other
terms in the low-$l$ regime. However, since we are interested in
the acoustic signatures of bispectrum induced by PMFs, i.e. the
moderate high-$l$ regime ($l\geq100$), we need not worry about the
momentum divergence.

 \begin{figure}[h]
    \centering
    \includegraphics[angle=-90,width=11.5cm]{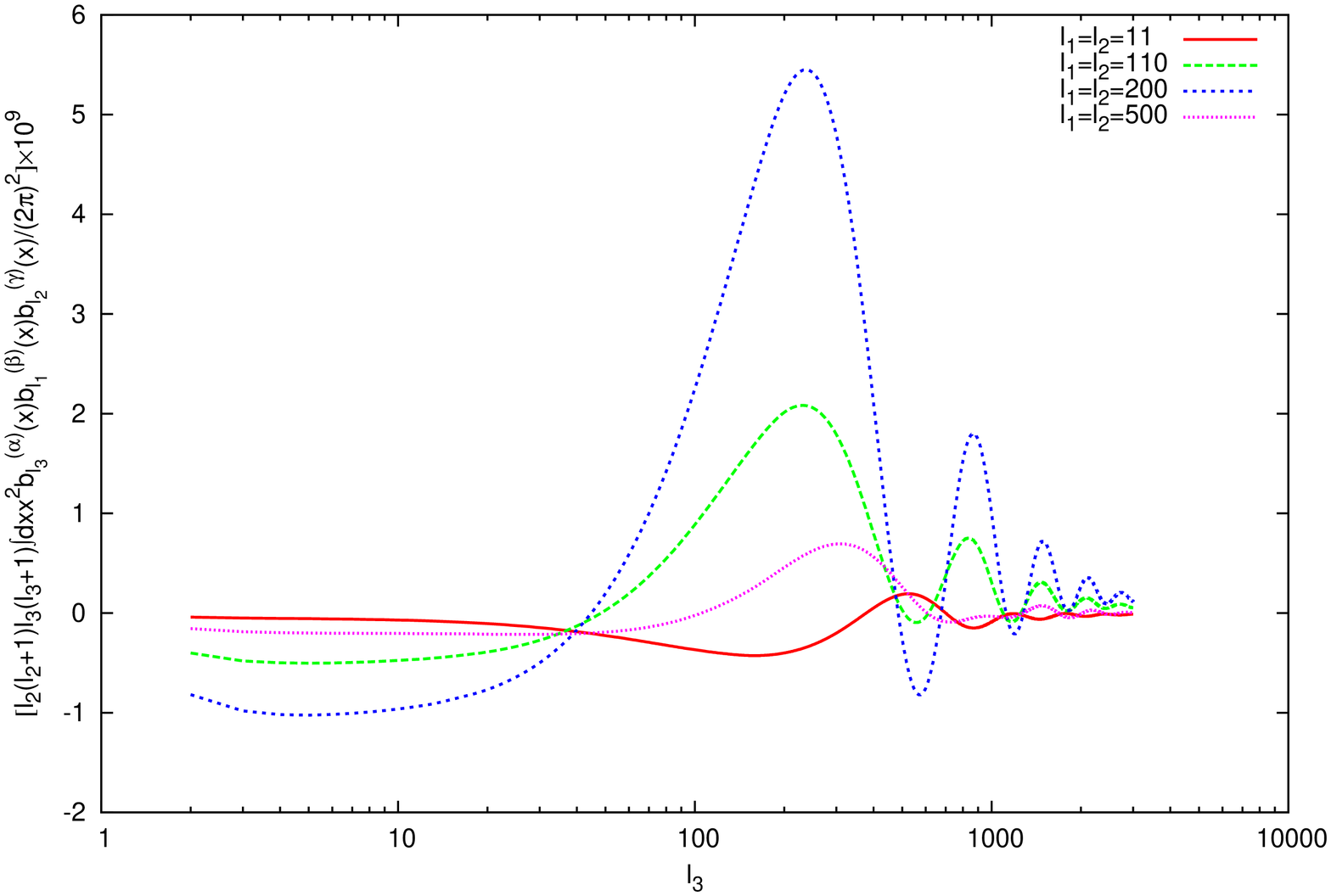}
    \caption{This figure shows the integral
$\Big[l_2(l_2+1)l_3(l_3+1)\int x^2dx~b_{l_3}^{(\a)}(x)
b_{l_1}^{(\b)}(x)b_{l_2}^{(\g)}(x) /(2\pi)^2\Big]\times10^{9}$ as
a function of $l_3$, with several parameter configurations
($l_1=l_2=11, 110, 200, 500$).}
    \label{lll1}
 \end{figure}
 \begin{figure}[h]
    \centering
    \includegraphics[angle=-90,width=11.5cm]{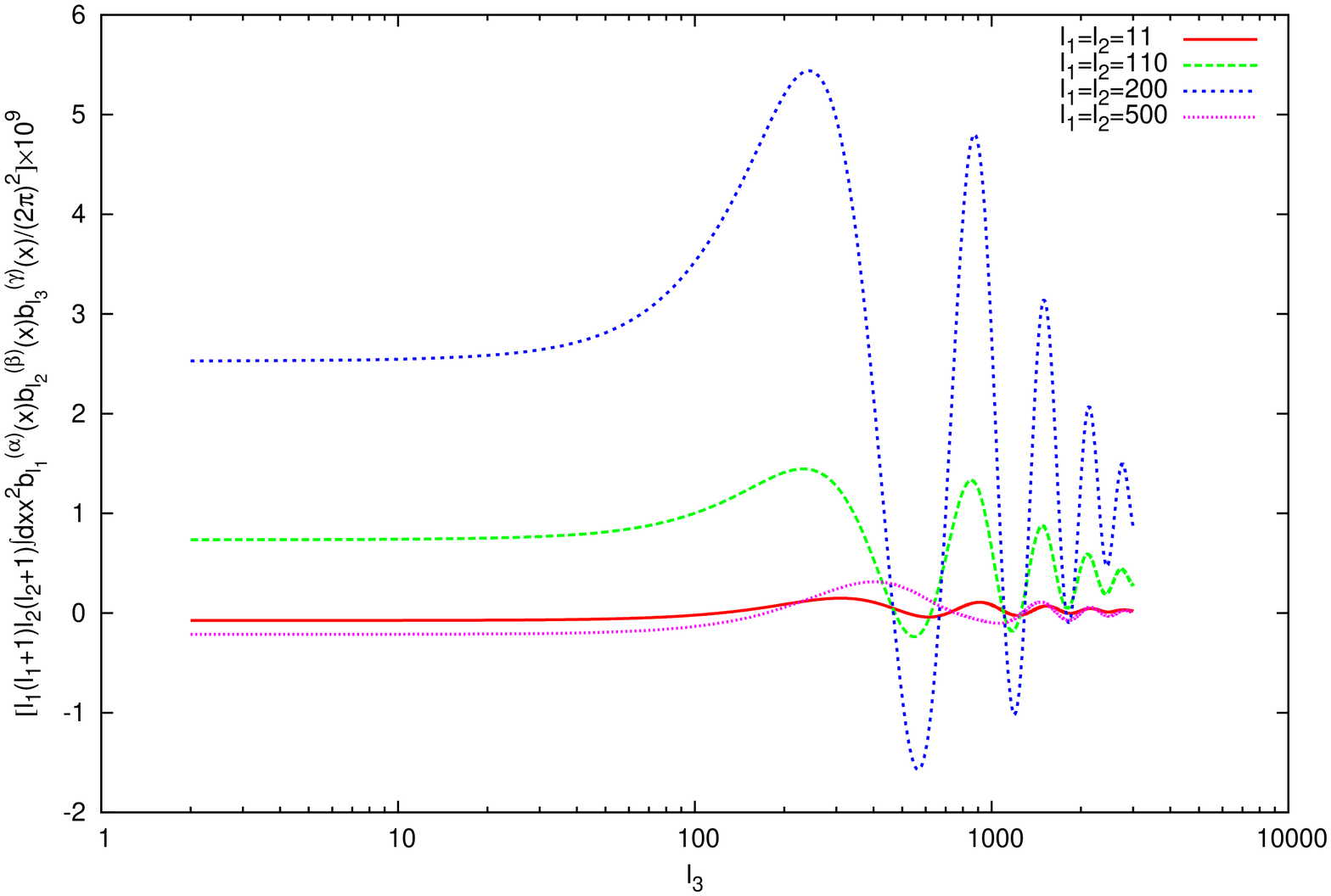}
    \caption{This figure shows the integral
$\Big[l_1(l_1+1)l_2(l_2+1)\int x^2dx~b_{l_1}^{(\a)}(x)
b_{l_2}^{(\b)}(x)b_{l_3}^{(\g)}(x) /(2\pi)^2\Big]\times10^{9}$ as
a function of $l_3$, with several parameter configurations
($l_1=l_2=11, 110, 200, 500$).}
    \label{3l1}
 \end{figure}
 \begin{figure}[h]
    \centering
    \includegraphics[angle=-90,width=11.5cm]{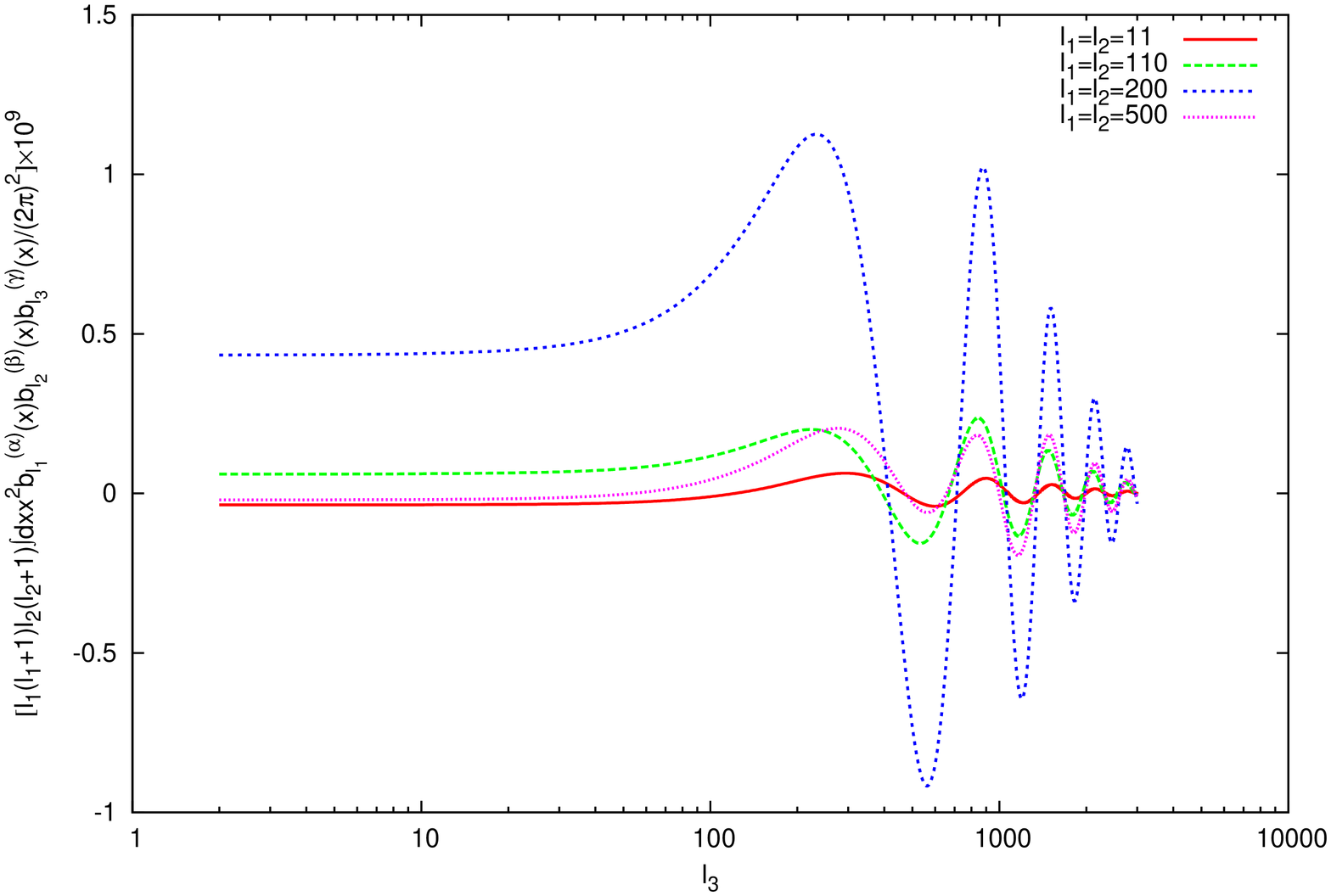}
    \caption{This figure shows the integral
$\Big[l_1(l_1+1)l_2(l_2+1)\int x^2dx~b_{l_1}^{(\a)}(x)
b_{l_2}^{(\b)}(x)b_{l_3}^{(\g)}(x) /(2\pi)^2\Big]\times10^{9}$ as
a function of $l_3$ in the case without considering the effect of
the Lorentz force. From this figure, we can see clearly that all
modes oscillate around zero in the limit of large $l_3$.}
    \label{nolorentz3l1}
 \end{figure}

Having the numerical results about the momentum integrations
(\ref{b_la})-(\ref{b_ld}), we can finally perform the
$x$-integrations to obtain the reduced bispectra. In Figure
(\ref{lll1}) and (\ref{3l1}), we plot the integrals
$\Big[l_2(l_2+1)l_3(l_3+1)\int x^2dx~b_{l_3}^{(\a)}(x)
b_{l_1}^{(\b)}(x)b_{l_2}^{(\g)}(x) /(2\pi)^2\Big]\times10^{9}$ and
$\Big[l_1(l_1+1)l_2(l_2+1)\int x^2dx~b_{l_1}^{(\a)}(x)
b_{l_2}^{(\b)}(x)b_{l_3}^{(\g)}(x) /(2\pi)^2\Big]\times10^{9}$
over $l_3$ by fixing $l_1=l_2=11, 110, 200, 500$, respectively. In
the numerical calculations we integrate $x$ from
$(\h_0-2\h_{\ast})$ to $(\h_0-0.1\h_{\ast})$, since the primary
signals come from the recombination epoch $\h_{\ast}$. From Figure
(\ref{lll1}) and (\ref{3l1}), we can see that in the Sachs-Wolfe
(SW) regime ($l\leq10$) our result presents a SW plateau, which is
consistent with that in \cite{Caprini:2009vk}. And in the high-$l$
regime, the integral shown in Figure (\ref{lll1}) is greatly
damped after the prominent first acoustic peak, since the phase of
$b^{(\g)}_l(x)$ oscillates rapidly as a function of $x$. In Figure
(\ref{3l1}) the integral also has a first acoustic peak, but damps
more slowly. And more importantly, from Figure (\ref{3l1}) we can
see clearly that the modes with different parameter sets
($l_1,l_2$) oscillate around different asymptotic values because
of the effect of the Lorentz force, which is exerted by PMFs on
the charged baryons. This feature is much different with the one
from the inflation scenario
\cite{Komatsu:2001rj,Babich:2004yc,Komatsu:2002db,Fergusson:2006pr}.
In order to illustrate this phenomenon more clearly, we perform
the above calculations without considering the Lorentz force term
in the baryon velocity equation (\ref{kbv}). Figure
(\ref{nolorentz3l1}) shows that, contrary to those with the
Lorentz force, all modes oscillate around zero in the limit of
large $l_3$.

 \begin{figure}[h]
    \centering
    \includegraphics[angle=-90,width=11.5cm]{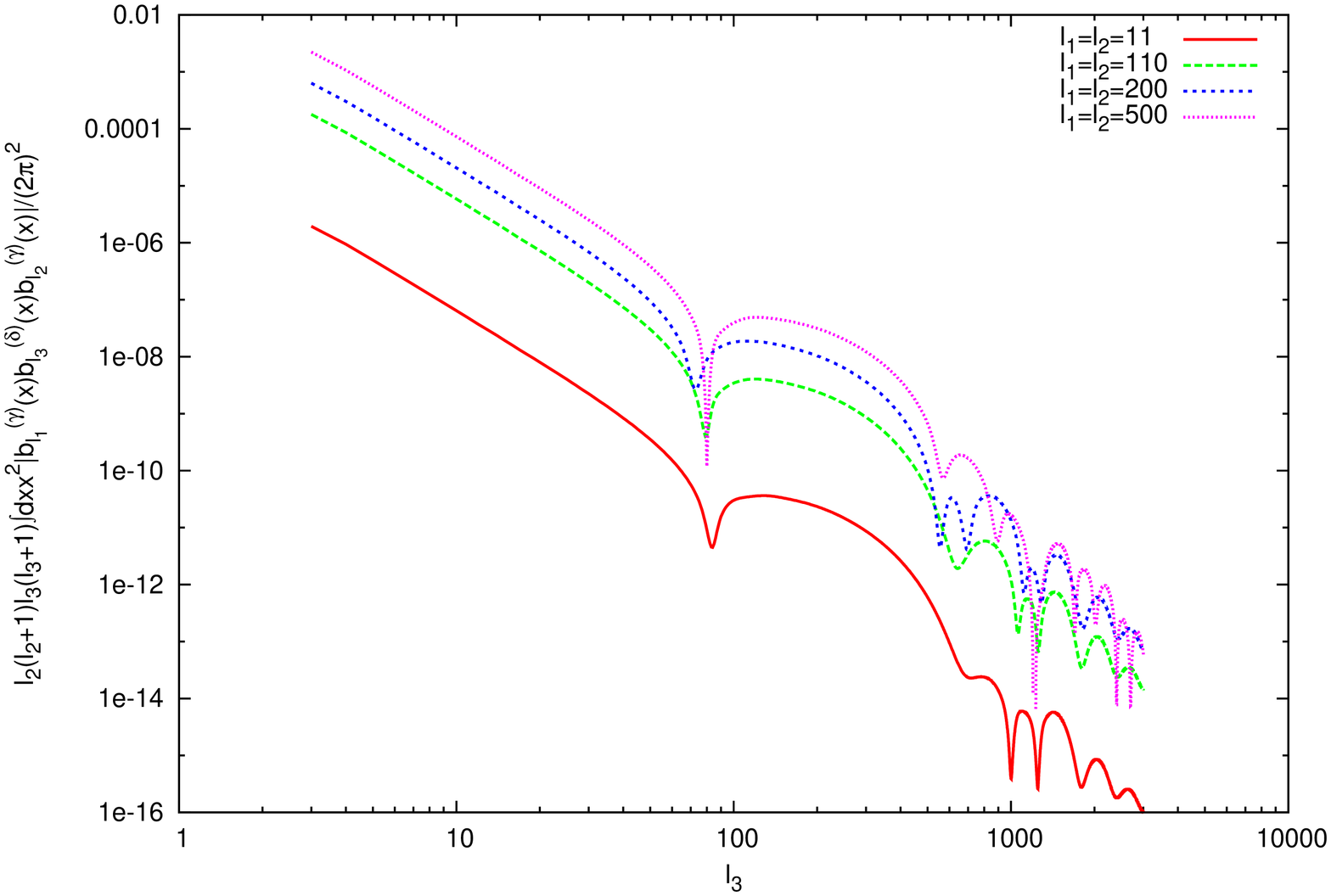}
    \caption{This figure shows the
absolute value of integral $\Big[l_2(l_2+1)l_3(l_3+1)\int
x^2dx~|b_{l_1}^{(\g)}(x) b_{l_3}^{(\d)}(x)b_{l_2}^{(\g)}(x)|
/(2\pi)^2\Big]$ as a function of $l_3$, with several parameter
configurations ($l_1=l_2=11, 110, 200, 500$).}
    \label{lll2}
 \end{figure}
 \begin{figure}[h]
    \centering
    \includegraphics[angle=-90,width=11.5cm]{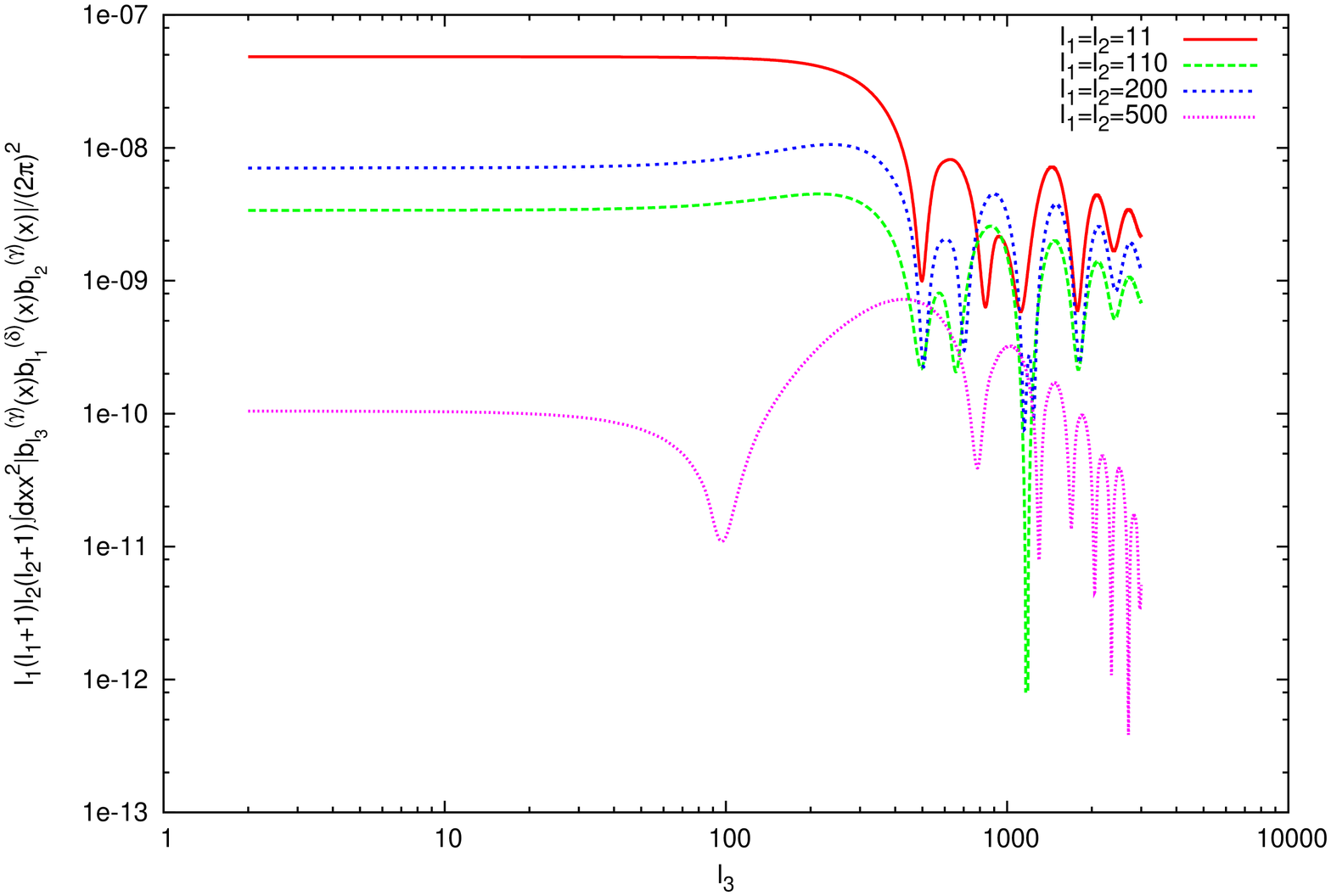}
    \caption{This figure show the
absolute value of the integral $\Big[l_1(l_1+1)l_2(l_2+1)\int
x^2dx~|b_{l_3}^{(\g)}(x) b_{l_1}^{(\d)}(x)b_{l_2}^{(\g)}(x)|
/(2\pi)^2\Big]$ as a function of $l_3$, with several parameter
configurations ($l_1=l_2=11, 110, 200, 500$).}
    \label{3l2}
 \end{figure}

In Figure (\ref{lll2}) and (\ref{3l2}), we plot the absolute
values of integrals $\Big[l_2(l_2+1)l_3(l_3+1)\int
x^2dx~|b_{l_1}^{(\g)}(x) b_{l_3}^{(\d)}(x)b_{l_2}^{(\g)}(x)|
/(2\pi)^2\Big]$ and $\Big[l_1(l_1+1)l_2(l_2+1)\int
x^2dx~|b_{l_3}^{(\g)}(x) b_{l_1}^{(\d)}(x)b_{l_2}^{(\g)}(x)|
/(2\pi)^2\Big]$ as a function of $l_3$ with $l_1=l_2=11, 110, 200,
500$. Since $b^{(2)}_{l_1l_2l_3}$ contains the factor
$b^{(\b)}_l$, its amplitude experiences a great suppression in the
high-$l$ regime as the same as $b^{(\b)}_l$. From Figure
(\ref{lll2}), we can see that the amplitude of first acoustic peak
is approximately of the same order as the one in Figure
(\ref{lll1}), however, in the very high-$l$ region ($l\geq1000$)
its amplitude damps in a power law form. In addition, from Figure
(\ref{bc}) and (\ref{ba}), we can see that the amplitude of
$b^{(\g)}_l$ is smaller than that of $b^{(\a)}_l$ by the order of
$\ma{O}(10^{-2})$. Furthermore, for the set of parameters
$(n_B=-2.9, B_{\lam}=9{\rm~nG})$, the coefficients in
(\ref{blll1})-(\ref{blll3}) are $\ma{N}_1\simeq
3.5\times10^{-12}$, $\ma{N}_2\simeq -6\times10^{-14}$ and
$\ma{N}_3\simeq -7\times10^{-18}$. After considering the
hierarchies between $\ma{N}_3$ and $\ma{N}_1$ (or $\ma{N}_2$), we
can safely neglect the $b^{(3)}_{l_1l_2l_3}$ term in our
calculations.

As a summary, from the above numerical calculations, we can see
that the typical amplitudes of the reduced bispectra
$l^4b^{(1)}_{l_1l_2l_3}/(2\pi)^2$ and
$l^4b^{(2)}_{l_1l_2l_3}/(2\pi)^2$ are of the order
$\ma{O}(10^{-20})$, which is comparable with the NG signals from
primary curvature perturbations
\cite{Komatsu:2001rj,Babich:2004yc,Komatsu:2002db,Fergusson:2006pr}.
In details, $b^{(1)}_{l_1l_2l_3}$ dominates the total reduced
bispectrum with a positive amplitude in the regime $l\geq100$. For
$b^{(2)}_{l_1l_2l_3}$, its amplitude in the high-$l$ regime, is of
the same order as the one of $b^{(2)}_{l_1l_2l_3}$, but the sign
is negative; while it diverges at the large scales.

\section{\label{concl}Conclusion}
Using the full radiation transfer function, we numerically
calculated the CMB angular bispectrum seeded by the compensated
magnetic scalar density modes. For PMFs characterized by the index
$n_B=-2.9$ and mean-field amplitude $B_{\lam}=9{\rm~nG}$, CMB
bispectrum is dominated by two primordial magnetic shapes. For the
reduced bispectrum $b^{(1)}_{l_1l_2l_3}$ (\ref{blll1}), which is
seeded by the ``local-type'' shape $f^{(1)}(k,q,p)$ (\ref{f1}),
both the profile and amplitude look similar to those of the
primary CMB anisotropies
\cite{Komatsu:2001rj,Babich:2004yc,Komatsu:2002db,Fergusson:2006pr},
(see Figure \ref{ba}, \ref{bc}, \ref{lll1} and \ref{3l1}).
However, for different parameter sets ($l_1,l_2$), such
``local-type'' estimator $b^{(1)}_{l_1l_2l_3}$ oscillates around
different asymptotic values in the high-$l_3$ regime because of
the effect of the Lorentz force, which is exerted by PMFs on the
charged baryons (see Figure \ref{3l1} and \ref{nolorentz3l1}).
This feature is  different from the one of inflation scenarios
where all modes approach to zero asymptotically in the high-$l$
limit. On the other hand, the second magnetic shape
$f^{(2)}(k,q,p)$ (\ref{f2}) appears only in the primordial
magnetic field model. However, the amplitude of the
Komatsu-Spergel estimator $b^{(2)}_{l_1l_2l_3}$ (\ref{blll2})
sourced by the shape $f^{(2)}(k,q,p)$ diverges in the low-$l$
regime because of the negative slope of shape. In the high-$l$
regime, this amplitude is approximately equal to that of the first
estimator $b^{(1)}_{l_1l_2l_3}$, but with a reversal phase.

In this paper we only calculated the magnetic angular bispectrum
with parameters ($n_B=-2.9$, $B_{\lam}=9{\rm~nG}$). In fact, the
results are strongly dependent on the magnetic parameters,
especially on the magnetic index $n_B$. Take $n_B=-2$ as an
example, the upper bound of comoving magnetic mean-field amplitude
becomes larger, $B_{\lam}\lesssim25{\rm~nG}$
\cite{Caprini:2009vk}. And more importantly, there will appear new
Gaussian and non-Gaussian features in the CMB anisotropies. At the
Gaussian level, the two-point function will have more powers on
the small scales, i.e. a blue tilt power spectrum; at the
non-Gaussian level, the bispectrum will no longer  diverge in the
IR limit. Finally, we would like to comment on the bispectrum from
the compensated magnetic anisotropic stress mode. Although the
amplitude of anisotropic stress $\pi^{(B)}_k$ is approximately
larger than that of density contrast $\D^{(B)}_k$ by a factor $3$,
$\pi^{(B)}_k$ mode only appears in the high order terms in the
tight-coupling expansion (\ref{ic_piB1})-(\ref{ic_piB9}). This
results in that the amplitudes of power spectra from $\pi^{(B)}_k$
mode are smaller than those from $\D^{(B)}_k$ mode, (see Figure
\ref{tt}, \ref{ee} and \ref{te}). So we can estimate that the
bispectrum from the compensated magnetic anisotropic stress should
be smaller than the one from the magnetic density.

\begin{acknowledgments}

BH thanks Kiyotomo Ichiki and Kazuhiko Kojima for the helpful
correspondence. RGC thanks the organizers and participants for
various discussions during the  workshop ``Dark Energy and
Fundamental Theory" held at Xidi, Anhui, China, May 28-June 6,
2010, supported by the Special Fund for Theoretical Physics from
the National Natural Science Foundation of China with grant No.
10947203, and the long-term workshop ``Gravity and Cosmology
2010", held at the Yukawa Institute for Theoretical Physics, Kyoto
University, Japan. This work was supported in part by the National
Natural Science Foundation of China under Grant Nos. 10535060,
10821504 and 10975168, and by the Ministry of Science and
Technology of China under Grant No. 2010CB833004.

\end{acknowledgments}

\appendix
\section{\label{cov}The covariant approach to cosmological perturbations}
In this Appendix we briefly review the covariant approach to
cosmological perturbations
\cite{Challinor:1998xk,Tsagas:2007yx,Hawking:1966qi,1971glc..conf....104E,
1973clp..conf....1E,King:1972td,Ellis:1989jt,Ellis:1989ju,Ellis:1990gi,
Stewart:1990fm,Hwang:1990am,Dunsby:1991xk,Bruni:1991xj,Bruni:1992dg,
Maartens:1995hh,Gebbie:1998fe,Maartens:1998xg,Challinor:1998aa,
Brechet:2009fa}.
Especially, the cosmological covariant formalisms with magnetic fields
are carefully studied in \cite{Tsagas:2007yx,
Tsagas:1997vf,Tsagas:1998jm,Tsagas:1999ft,
Tsagas:1999tu,Tsagas:2004kv,Barrow:2006ch}.
In order to determine the time direction, we
define a unit timelike 4-velocity vector tangent to the worldline
of the observer
 \be\label{4vel}
 u^a=\f{dx^a}{d\tau}\;,\qquad u_au^a=-1\;,\ee
where $\tau$ is the proper time of the fundamental observer. Then
we introduce an orthogonal tensor $h_{ab}=g_{ab}+u_au_b$ with
respect to $u_a$
 to define the space direction
at each spacetime point. Using the vector field $u_a$ and
projector tensor $h_{ab}$, we can decompose any spacetime quantity
into its irreducible temporal and spatial parts. Moreover, we can
also use these fields to define the covariant time and spatial
derivatives of any tensor field $S_{ab\cdots}^{~~~~cd\cdots}$
according to
 \be\label{deriv}
 \dot S_{ab\cdots}^{~~~~cd\cdots}=u^e\9_eS_{ab\cdots}^{~~~~cd\cdots}\;,\qquad
 D_eS_{ab\cdots}^{~~~~cd\cdots}=h_e^{~s}h_a^{~f}h_b^{~p}h_q^{~c}h_r^{~d}\cdots
 \9_sS_{fp\cdots}^{~~~~qr\cdots}\;,
 \ee
respectively.

In this paper we use the convention about the effective volume element
$\ve_{abc}$ and the spacetime volume element $\h_{abcd}$ as
 \be\label{eps}
 \ve_{abc}=\h_{abcd}u^d\;,\ee
where the totally skew pseudotensor is defined as
$\h^{0123}=[-\det(g_{ab})]^{-1/2}$. Furthermore, $\h^{abcd}$ is
parallelly transported ($\h^{abcd}_{~~~~;e}=0$) and satisfies some
 algebras as
 \bea\label{eta algeb}
 \h^{abcd}\h_{efgh}&=&-4!\d^{[a}_e\d^b_f\d^c_g\d^{h]}_d\;,\\
 \h^{abcs}\h_{efgs}&=&-3!\d^{[a}_e\d^b_f\d^{c]}_g\;,\\
 \h^{abts}\h_{efts}&=&-4\d^{[a}_e\d^{b]}_f\;,\\
 \h^{arts}\h_{erts}&=&-3!\d^a_e\;,\\
 \h^{prts}\h_{prts}&=&-4!\;,\eea
where the square bracket in the superscript represents for the
antisymmetric part of corresponding tensors.

In General Relativity, the local gravitational interaction is
described by Ricci tensor $R_{ab}$, while the non-local long-range
interaction, such as gravitational waves or tidal forces, is
determined by  Weyl conformal curvature tensor $C_{abcd}$. The
decomposition of the gravitational field into its local and
non-local parts is given by
 \be
 R_{abcd}=C_{abcd}+\f{1}{2}(g_{ac}R_{bd}+g_{bd}R_{ac}-g_{bc}R_{ad}-g_{ad}R_{bc})
 -\f{1}{6}R(g_{ac}g_{bd}-g_{ad}g_{bc})\;,\ee
where  Weyl tensor shares all the symmetries of  Riemann tensor
and it is trace-free $C^c_{~acb}=0$. Furthermore, we can define
the irreducible electric and magnetic parts of the Weyl tensor
 \be
 E_{ab}=C_{acbd}u^cu^d\;,\qquad H_{ab}=\f{1}{2}\ve_a^{~cd}C_{cdbe}u^e\;.\ee
Then the Weyl tensor can be rewritten by these two tensors
 \be
 C_{abcd}=(g_{abqp}g_{cdsr}-\h_{abqp}\h_{cdsr})u^qu^sE^{pr}
 -(\h_{abqp}g_{cdsr}+g_{abqp}\h_{cdsr})u^qu^sH^{pr}\;,\ee
where $g_{abcd}=g_{ac}g_{bd}-g_{ad}g_{bc}$.

The energy-momentum
tensor of a general (imperfect) fluid defined by the observer $u_a$
can be decomposed into
 \be\label{energymomen}
 T_{ab}=\r u_au_b+ph_{ab}+2q_{(a}u_{b)}+\pi_{ab}\;,\ee
where $\r=T_{ab}u^au^b$, $p=T_{ab}h^{ab}/3$, $q_a=-h_a^{~b}T_{bc}u^c$
and $\pi_{ab}=h_{\la a}^{~~c}h_{b\ra}^{~~d}T_{cd}=
h_{(a}^{~~c}h_{b)}^{~~d}T_{cd}-\f{1}{3}h^{cd}T_{cd}h_{ab}$
are the energy density,
isotropic pressure, energy-flux and anisotropic stress
tensor of the fluid, respectively.

In order to characterize the observer's motion we need to
decompose the 4-velocity gradient into the following irreducible
kinematical quantities relative to the $u_a$-congruence
 \be\label{velgrad}
 \9_bu_a=\s_{ab}+\o_{ab}+\f{1}{3}\Theta h_{ab}-A_au_b\;,\ee
where $\s_{ab}=D_{\la b}u_{a\ra}$, $\o_{ab}=D_{[b}u_{a]}$,
$\Theta=\9_au^a=D_au^a$ and $A_a=\dot u_a=u^b\9_bu_a$ are
respectively the shear and vorticity tensors, the volume expansion
scalar, and the 4-acceleration vector. In addition, it is useful
to define the vorticity vector $\o_a=\ve_{abc}\o^{bc}/2$ (with
$\o_{ab}=\ve_{abc}\o^c$) instead of the vorticity tensor.

\subsection{Linearized Einstein equations}
 Dynamical equations:
 \bea
 \dot\r+(\r+p)\Theta+D^aq_a&=&0\;,\label{den}\\
 \dot\Theta+\f{1}{3}\Theta^2+\f{1}{2}(\r+3p)-D^aA_a&=&0\;,\label{ray}\\
 \dot q_a+\f{4}{3}\Theta
 q_a+(\r+p)A_a+D_ap+D^b\pi_{ab}&=&0\;,\label{mom}\\
 \dot\o_{\la a\ra}+\f{2}{3}\Theta\o_a+\f{1}{2}{\rm
 curl}~A_a&=&0\;,\label{vor}\\
 \dot\s_{\la
 ab\ra}+\f{2}{3}\Theta\s_{ab}+E_{ab}-\f{1}{2}\pi_{ab}-D_{\la
 a}A_{b\ra}&=&0\;,\label{shear}\\
 \dot E_{\la ab\ra}+\Theta E_{ab}-{\rm
 curl}~H_{ab}+\f{1}{2}(\r+p)\s_{ab}&&\nn\\
 +\f{1}{2}\dot\pi_{\la
 ab\ra}+\f{1}{2}D_{\la a}q_{b\ra}+\f{1}{6}\Theta\pi_{ab}&=&0\;,\label{elec}\\
 \dot H_{\la ab\ra}+\Theta H_{ab}+{\rm curl}~E_{ab}-\half{\rm
 curl}~\pi_{ab}&=&0\;.\label{mag}\eea
Constraint equations:
 \bea
 D_a\o^a&=&0\;,\label{vor2}\\
 D^b\s_{ab}-{\rm curl}~\o_a-\f{2}{3}D_a\Theta+q_a&=&0\;,\label{shear2}\\
 {\rm curl}~\s_{ab}+D_{\la a}\o_{b\ra}-H_{ab}&=&0\;,\label{shear3}\\
 D^bE_{ab}+\half D^b\pi_{ab}-\f{1}{3}D_a\r+\f{1}{3}\Theta
 q_a&=&0\;,\label{elec2}\\
 D^bH_{ab}+\half {\rm curl}~q_a-(\r+p)\o_a&=&0\;.\label{mag2}\eea

\subsection{Two key variables}
It is convenient to define two key variables in the covariant
approach
 \be
 \D^{(i)}_a=\f{a}{\r^{(i)}}D_a\r^{(i)}\;,\qquad \ma{Z}_a=aD_a\Theta\;,\ee
where $i=\g, \n, b, c$.
Taking the spatial gradient of the density evolution
equation (\ref{den}), we arrive at
 \be\label{Da} \r^{(i)}\dot\D^{(i)}_a+(\r^{(i)}+p^{(i)})(\ma{Z}_a+a\Theta
 A_a)+aD_aD^bq^{(i)}_b+a\Theta D_ap^{(i)}-p^{(i)}\Theta\D_a=0\;,\ee
For $\ma{Z}_a$, by virtue of the Raychaudhuri
equation (\ref{ray}) we have
 \be\label{maZ}
 \dot\ma{Z}_a+\f{2\Theta}{3}\ma{Z}_a+\half\r\D_a+\f{3}{2}aD_ap
 +a\le[\f{1}{3}\Theta^2+\half(\r+3p)\ri]A_a-aD_aD^bA_b=0\;.\ee

\section{\label{linear_equations}Equations for matter components}
Under the ideal MHD approximation, the energy-momentum tensors for
the five matter components are
 \be\label{energymomen2}
 T^{(i)}_{ab}=\r^{(i)} u_au_b+p^{(i)}h_{ab}+2q^{(i)}_{(a}u_{b)}+\pi^{(i)}_{ab}\;,\ee
with $i=\g, \n, b, c$, and
 \be\label{magTab2}
 T_{ab}^{(B)}=\f{1}{4\pi}\le[\f{1}{2}B^2u_au_b+\f{1}{6}B^2h_{ab}\ri]+\pi^{(B)}_{ab}\;.\ee
Since the total energy-momentum tensor is conserved $\9^bT_{ab}=0$, for
each component we have
 \be \9^bT^{(i)}_{ab}=J^{(i)}_a=E^{(i)}u_a+M^{(i)}_a\;,
 \quad \sum_i J^{(i)}_a=0\;,\ee
where $E^{(i)}$ is the energy transfer and $M^{(i)}_a$ the
momentum transfer for the $i$-species. For simplicity, in this
work we assume the energy transfer vanishes at the linear order
($E^{(i)}\sim 0$), this gives the energy conservation for each
matter component
 \be \dot\r^{(i)}+\Theta(\r^{(i)}+p^{(i)})+D^aq_a^{(i)}=0\;.\ee

\subsection{Photons}
For photons, it is convenient to expand the total intensity brightness ${\bf I}(E,e^c)$
 in terms of the spherical multipole
 \be\label{intensity}
 {\bf I}(E,e^c)=\sum_{l=0}^{\infty}I_{A_l}(E)e^{A_l}\;,\ee
where $e^c$ is a unit spacelike vector orthogonal to $u_a$. For CMB,
it is usual to define the bolometric multipoles by integrating over
energy without loss of information
 \be\label{multipole}
 I_{A_l}=\D_l\int_0^{\infty} I_{A_l}(E) dE\;,\ee
with
 \be \D_l=\f{4\pi 2^l(l!)^2}{(2l+1)!}\;.\ee
The first three multipoles are respectively
 \be I=\r^{(\gamma)}\;,\qquad I_a=q^{(\g)}_a\;,\qquad
 I_{ab}=\pi_{ab}^{(\gamma)}\;.\ee

The Boltzmann hierarchies for the total intensity of photons are
 \bea\label{boltz1}
 &&\dot I_{A_l}+\f{4}{3}\T I_{A_l}+D^bI_{bA_l}+\f{l}{(2l+1)}D_{\la
 a_l}I_{A_{l-1}\ra}+\f{4}{3}IA_{a_1}\d_{l1}+\f{8}{15}I\s_{a_1a_2}\d_{l2}\nn\\
 &&=-n_e\s_T\le[I_{A_l}-I\d_{l0}-\f{4}{3}Iv_{a_1}\d_{l1}-\f{1}{10}I_{a_1a_2}\d_{l2}\ri]\;,\eea
where the right hand side terms stand for the Thompson scattering.
The first three multipole hierarchy equations are listed as
follows.

Monopole ($l=0$):
 \be\label{gamma l=0} \dot\r^{(\gamma)}+\f{4}{3}\Theta\r^{(\gamma)}+D^aq_a^{(\gamma)}=0\;,\ee
Usually one uses the spatial gradient of energy conservation
equation, instead of (\ref{gamma l=0}),
 \be\label{gamma Da} \dot\D_a^{(\gamma)}+\f{4}{3}(\ma{Z}_a+a\Theta
 A_a)+\f{a}{\r^{(\g)}}D_aD^bq^{(\g)}_b=0\;.\ee

Dipole ($l=1$):
 \be\label{gamma l=1} \dot q^{(\gamma)}_a+\f{4}{3}\Theta q^{(\gamma)}_a+\f{1}{3}D_a\r^{(\gamma)}
 +D^b\pi^{(\gamma)}_{ab}+\f{4}{3}\r^{(\gamma)}A_a=n_e\s_T\le[\f{4}{3}\r^{(\gamma)}v^{(b)}_a-q^{(\gamma)}_a\ri]\;.\ee

Quadrupole ($l=2$):
 \be\label{gamma l=2}
 \dot\pi^{(\gamma)}_{ab}+\f{4}{3}\Theta\pi^{(\gamma)}_{ab}+D^cI_{abc}+\f{2}{5}D_{\la
 b}q^{(\gamma)}_{a\ra}+\f{8}{15}\r^{(\gamma)}\s_{ab}=-\f{9}{10}n_e\s_T\pi^{(\gamma)}_{ab}\;.\ee
Note that the monopole equation (\ref{gamma l=0}) is equivalent to
the equation of energy conservation of the photons, and the dipole
one (\ref{gamma l=1}) to the momentum conservation equation with
the Thompson scattering.

\subsection{Massless neutrinos}

For massless neutrinos, the Boltzmann hierarchies are similar with
the ones of photons,  except that the latter has the Thompson
collision term,
 \bea\label{nuboltz1}
 &&\dot G_{A_l}+\f{4}{3}\T G_{A_l}+D^bG_{bA_l}+\f{l}{(2l+1)}D_{\la
 a_l}G_{A_{l-1}\ra}+\nn\\
 &&\f{4}{3}GA_{a_1}\d_{l1}+\f{8}{15}G\s_{a_1a_2}\d_{l2}
 =0\;.\eea
The monopole, dipole and quadrapole equations are, respectively,

monopole ($l=0$)
 \be\label{nu l=0} \dot\r^{(\n)}+\f{4}{3}\Theta\r^{(\n)}+D^aq_a^{(\n)}=0\;,\ee
or replaced by
 \be\label{nu Da} \dot\D_a^{(\n)}+\f{4}{3}(\ma{Z}_a+a\Theta
 A_a)+\f{a}{\r^{(\n)}}D_aD^bq^{(\n)}_b=0\;.\ee

dipole ($l=1$)
 \be\label{nu l=1} \dot q^{(\n)}_a+\f{4}{3}\Theta
 q^{(\n)}_a+D_ap^{(\n)}+\f{1}{3}D^b\pi^{(\n)}_{ab}+\f{4}{3}\r^{(\n)}A_a=0\;.\ee

quadrupole ($l=2$)
 \be\label{nu l=2}
 \dot\pi^{(\n)}_{ab}+\f{4}{3}\Theta\pi^{(\n)}_{ab}+D^cG_{abc}+\f{2}{5}D_{\la
 b}q^{(\n)}_{a\ra}+\f{8}{15}\r^{(\n)}\s_{ab}=0\;.\ee

\subsection{Baryons}
For baryons, the energy conservation equation gives
 \be\label{brho}
 \dot\r^{(b)}+(\r^{(b)}+p^{(b)})\Theta+(\r^{(b)}+p^{(b)})D^av^{(b)}_a=0\;,\ee
and by taking the spatial gradient of (\ref{brho}), we obtain
 \be\label{bDa} \r^{(b)}\dot\D_a^{(b)}+(\r^{(b)}+p^{(b)})(\ma{Z}_a+a\Theta
 A_a+aD_aD^cv_c^{(b)})+a\Theta D_ap^{(b)}-p^{(b)}\Theta\D_a^{(b)}=0\;.\ee
The momentum conservation reads
 \bea\label{bv} \le(\r^{(b)}+p^{(b)}\ri)\le(\dot
 v^{(b)}_a+A_a\ri)+\dot p^{(b)}v^{(b)}_a+\f{1}{3}(\r^{(b)}+p^{(b)})\Theta v^{(b)}_a+D_ap^{(b)}
 =\nn\\
 -\le[n_e\s_T\le(\f{4}{3}\r^{(\gamma)}v^{(b)}_a-q^{(\gamma)}_a\ri)
 +D^b\pi^{(B)}_{ab}+D_ap^{(B)}\ri]\;,\eea
where the momentum transfer for baryons is due to the Thompson
scattering and Lorentz force from PMFs.

\subsection{Cold Dark Matter}
Since CDM only gravitates, the density contrast equation is given
by
 \be\label{crho} \dot\r^{(c)}+\Theta\r^{(c)}+\r^{(c)}D^av_a^{(c)}=0\;,\ee
or
 \be\label{cDa} \dot\D_a^{(c)}+\ma{Z}_a+a\Theta
 A_a+aD_aD^bv_b^{(c)}=0\;,\ee
and the velocity equation is
 \be\label{cv} \dot v^{(c)}_a+\f{1}{3}\Theta v^{(c)}_a+A_a=0\;.\ee
They are simple.

\end{document}